\documentclass{aastex6}

\begin{document}


\title{Consistent SPH Simulations of Protostellar Collapse and Fragmentation}


\author{Ruslan Gabbasov}
\affil{Instituto de Ciencias B\'asicas e Ingenier\'{\i}as, Universidad Aut\'onoma del
Estado de Hidalgo (UAEH),\\
Ciudad Universitaria, Carretera Pachuca-Tulacingo km. 4.5 S/N, Colonia Carboneras,\\
Mineral de la Reforma, C.P. 42184, Hidalgo, Mexico}

\author{Leonardo Di G. Sigalotti, Fidel Cruz}
\affil{\'Area de F\'{\i}sica de Procesos Irreversibles, Departamento de Ciencias B\'asicas,\\
Universidad Aut\'onoma Metropolitana-Azcapotzalco (UAM-A), Av. San Pablo 180,\\
C.P. 02200, Ciudad de M\'exico, Mexico}

\author{Jaime Klapp\altaffilmark{1}}
\affil{Departamento de F\'{\i}sica, Instituto Nacional de Investigaciones Nucleares (ININ),\\
Carretera M\'exico-Toluca km. 36.5, La Marquesa, 52750 Ocoyoacac, Estado de M\'exico, Mexico}
\author{Jos\'e M. Ram\'{\i}rez-Velasquez\altaffilmark{1}}
\affil{Centro de F\'{\i}sica, Instituto Venezolano de Investigaciones Cient\'{\i}ficas (IVIC),\\
Apartado Postal 20632, Caracas 1020A, Venezuela}

\altaffiltext{1}{ABACUS-Centro de Matem\'aticas Aplicadas y C\'omputo de Alto Rendimiento,\\
Departamento de Matem\'aticas, Centro de Investigaci\'on y de Estudios Avanzados (Cinvestav-IPN),\\
Carretera M\'exico-Toluca km. 38.5, La Marquesa, 52740 Ocoyoacac, Estado de M\'exico, Mexico}


\begin{abstract}

We study the consistency and convergence of smoothed particle hydrodynamics (SPH), as a
function of the interpolation parameters, namely the number of particles $N$, the
number of neighbors $n$, and the smoothing length $h$, using simulations
of the collapse and fragmentation of protostellar rotating cores. The calculations are
made using a modified version of the GADGET-2 code that employs an improved scheme for
the artificial viscosity and power-law dependences of $n$ and $h$ on $N$, as was recently
proposed by Zhu et al., which comply with the combined limit $N\to\infty$,
$h\to 0$, and $n\to\infty$ with $n/N\to 0$ for full SPH consistency, as the domain
resolution is increased. We apply this realization to the
``standard isothermal test case'' in the variant calculated by Burkert \& Bodenheimer
and the Gaussian cloud model of Boss to investigate the response of the method to adaptive
smoothing lengths in the presence of large density and pressure gradients. The degree
of consistency is measured by tracking how well the estimates of the consistency integral
relations reproduce their continuous counterparts. In particular, $C^{0}$ and $C^{1}$
particle consistency is demonstrated, meaning that the calculations are close to
second-order accuracy. {\bf As long as $n$ is increased with $N$, mass resolution also
improves as the minimum resolvable mass $M_{\rm min}\sim n^{-1}$. This aspect allows
proper calculation of small-scale structures in the flow associated with the formation
and instability of protostellar disks around the growing fragments, which are seen to
develop a spiral structure and fragment into close binary/multiple systems as supported
by recent observations}.

\end{abstract}



\section{Introduction} \label{sec:intro}

The method of smoothed particle hydrodynamics (SPH) was developed in the late 1970s
by \cite{Gingold77} and \cite{Lucy77} as a numerical tool for solving the equations
of gravitohydrodynamics in three-dimensional open space. Today, the use of SPH spans
many areas of astrophysics and cosmology as well as a broad range of fluid and
solid mechanics related areas. However, despite its extensive applications and 
recent progress in consolidating its theoretical foundations, SPH still has unknown
properties that need to be investigated. A fundamental numerical aspect of SPH is
the lack of particle consistency, which affects the accuracy and convergence of the
method. Several modified techniques and corrective methods have been proposed to
restore particle consistency in fluid dynamics calculations \citep{Li96,Bonet99,Chen99,
Liu03,Zhang04,Liu06,Sibilla15}; the most successful ones being those based on Taylor series
expansions of the kernel approximations of a function and its derivatives. If $m$
derivatives are retained in the series expansions, the resulting kernel and particle
approximations will have ($m+1$)th-order accuracy or $C^{m}$ consistency. However,
the improved accuracy of these methods comes at the price of involving matrix
inversions, which represent a major computational burden for time-evolving simulations
and eventually a loss of numerical stability due to matrix conditioning
for some specific problems. On the other hand, while these corrective methods solve
for particle inconsistency due to truncation of the kernel at model boundaries, it is
not clear how irregular particle distributions and the use of variable 
smoothing lengths affect the consistency (and therefore the accuracy) of the solutions.
Recently, \cite{Litvinov15} showed that the condition for the particle approximation
to restore $C^{0}$ consistency and achieve asymptotic error decay is that the
volumes defined by the particles and the inter-particle faces partition the entire
domain, i.e., constitute a partition of unity. They found that this condition is
satisfied by relaxing the particles under a constant pressure field by keeping the
particle volumes invariant, yielding convergence rates for such a relaxed distribution 
that are the same as those for particles on a perfect regular lattice. Quite curiously, 
they also observed that the relaxed particle distributions obtained this way resemble
that of liquid molecules resulting from microscopic simulations. A method to improve
the SPH estimate of derivatives which is not affected by particle disorder was also
devised recently by \cite{Sibilla15}.

In comparison little work has been done to improve the SPH consistency in astrophysical
applications. In many cases, especially those involving self-gravitating flows, large
density gradients arise and an adaptive kernel is used to guarantee spatial resolution
in regions of high density. It has long been recognized that spatially adaptive 
calculations where a variable smoothing length is employed turn out to be inconsistent
\citep{Liu06}. It was not until recently that \cite{Zhu15} identified another source
of particle inconsistency associated with the finite number of neighbors within the
compact support of a smoothed function. It is common practice in SPH calculations to
assume that a large number of total particles, $N$, and a small smoothing length, $h$,
are sufficient conditions to achieve consistent solutions, while holding the number of
neighbor particles, $n$, fixed at some value $n\ll N$. \cite{Zhu15} demonstrated that
$C^{0}$ particle consistency, i.e., satisfaction of the discrete normalization condition of
the kernel function can only be achieved when $n$ is sufficiently large for which the finite
SPH sum approximation approaches the continuous limit. This result is consistent with the
error analysis of the SPH representation of the continuity and momentum equations 
carried out by \cite{Read10}, who found that particle consistency is completely lost
due to zeroth-order error terms that would persist when working with a finite number of
neighbors even though $N\to\infty$ and $h\to 0$. Indeed, as the resolution is increased,
approaching the limit $N\to\infty$ and $h\to 0$, the overall error will grow at a
faster rate if the magnitude of the zeroth-order error terms remains
constant. Based on these observations, full particle consistency is possible in SPH
only if the joint limit $N\to\infty$, $h\to 0$, and $n\to\infty$ is satisfied
\citep{Zhu15}. However, we recall that this combined limit was first noted by
\cite{Rasio00} using a simple linear analysis on one-dimensional sound wave propagation.
In particular, he found that SPH is fully consistent in this limit with $N\to\infty$
faster than $n$ such that $n/N\to 0$.

On the other hand, \cite{Monaghan92} conjectured that for quasi-regularly distributed
particles, the discretization error made when passing from the continuous kernel to the
particle approximation is proportional to $(\log n)^{d}/n$, where $d$ is the dimension.
For $n\gg 1$, \cite{Zhu15} parameterized this error
as $\sim n^{-\gamma}$, where $\gamma$ varies from 0.5 for a random
distribution to 1 for a perfectly regular lattice of particles. Combining this with
the leading error ($\propto h^{2}$) of the continuous kernel approximation for most
commonly used kernel forms, \cite{Zhu15} derived the scaling relations $n\propto N^{1/2}$
and $h\propto N^{-1/6}$, which satisfy the joint limit as the domain resolution is 
progressively increased. A recent analysis on standard SPH has demonstrated that using
the above scalings $C^{0}$ consistency is fully restored for both the estimates of the
function and its derivatives in contrast to the case where $n$ is fixed to a
constant small value, with the numerical solution becoming also insensitive to the degree of 
particle disorder \citep{Sigalotti16}.

While these results are promising, it remains to investigate the response of the method
for spatially adaptive calculations in the presence of large gradients where the loss of
particle consistency is known to be most extreme. In particular, most of the above
analyses are based on static convergence tests for analytical functions in two- or
three-space dimensions using either uniformly or irregularly distributed point sets,
or on dynamical test problems for which an analytical solution is known in advance, and
therefore the results obtained are limited to idealized circumstances. As was emphasized
by \cite{Zhu15}, the lack of consistency associated with particle disorder and spatial
adaptivity is not specific to a particular SPH scheme but is rather a generic problem.
It would therefore be desirable to test the present method for more complex models
as those involving the solution of the equations of hydrodynamics coupled to
gravity in three-space dimensions. To do so we choose as a problem the gravitational
collapse and fragmentation of an initially rotating protostellar cloud, using a
modified version of the GADGET-2 code \citep{Springel05}. As templates for the model
clouds we use the ``standard isothermal test case'' in the variant calculated by
\cite{Burkert93} and the centrally condensed, Gaussian cloud model of \cite{Boss91}
coupled to a barotropic equation of state to mimic the nonisothermal collapse. The
simulations will then allow to better understand the impact of varying the number of
neighbors as the resolution is increased on the SPH discretization errors, which
will naturally emerge from the density estimate itself and the SPH momentum equation.
The convergence and accuracy of the simulations is measured by evaluating how well
the particle approximation of the integral consistency relations (or moments of the
kernel) are satisfied during the evolution. 

{\bf A further implication of the consistency scaling relations on protostellar collapse
calculations is the improved mass resolution. Since the minimum resolvable mass,
$M_{\rm min}$, scales with $h$ as $h^{3}$, this implies that $M_{\rm min}\sim n^{-1}$.
Although the collapse models proposed here start from ideal conditions, this aspect
has an important impact on the outcome of the simulations, where well-defined, rotating
circumstellar disks are seen to form around the growing fragments, which then increase in mass,
develop spiral arms, and fragment to produce small-scale binary/multiple protostellar
systems. This result is consistent with recent observations of L1448 IRS3B \citep{Tobin16}:
a close triple protostar system where two of the protostars formed by fragmentation of
a massive disk with a spiral structure surrounding a primary, young star formed from
the collapse of a larger cloud of gas and dust. While based principally on the relative
proximity of the companion stars, this observation provides for the first time direct 
evidence of protostellar disk fragmentation as a mechanism for the formation of 
close binary/multiple young stars}. 

\section{The issue of consistency} \label{sec:consistency}

We start by recalling that the kernel (or smoothed) estimate of a scalar function 
$f({\bf r})$, where $f$ may be either the density, $\rho$, or the gas pressure, $p$,
is defined by
\begin{equation}
\langle f({\bf r})\rangle =\int _{{\cal R}^{3}}f({\bf r}^{\prime})
W(|{\bf r}-{\bf r}^{\prime}|,h)d^{3}{\bf r}^{\prime},
\end{equation}
where the volume integration is taken over the whole real space, 
${\bf r}=(x,y,z)$ denotes position, and $W$ is the kernel interpolation function,
which must be positive definite, symmetric, monotonically decreasing, and satisfy the 
normalization condition
\begin{equation}
\int _{{\cal R}^{3}}W(|{\bf r}-{\bf r}^{\prime}|,h)d^{3}{\bf r}^{\prime}=1,
\end{equation}
together with the Dirac-$\delta$ function property that is observed when $h\to 0$. 
Moreover, suitable kernels should have a compact support so that 
$W(|{\bf r}-{\bf r}^{\prime}|,h)=0$ for $|{\bf r}-{\bf r}^{\prime}|\geq kh$, where $k$
is some integer that depends on the kernel function itself. Making the change of
variable $|{\bf r}-{\bf r}^{\prime}|\to h|{\bf r}-{\bf r}^{\prime}|$, it is easy to
show that the following scaling relation holds \citep{Sigalotti16}
\begin{equation}
W(h|{\bf r}-{\bf r}^{\prime}|,h)=\frac{1}{h^{\nu}}W(|{\bf r}-{\bf r}^{\prime}|,1),
\end{equation}
for any SPH kernel function, where $\nu =1$, 2, and 3 in one-, two-, and three-space
dimensions, respectively. 

If we expand in Taylor series $f({\bf r}^{\prime})$ around ${\bf r}^{\prime}={\bf r}$,
make ${\bf r}\to h{\bf r}$ and ${\bf r}^{\prime}\to h{\bf r}^{\prime}$, use Eq. (3), and 
insert the result in the kernel approximation (1), we obtain for the function estimate
the relation
\begin{equation}
\langle f(h{\bf r})\rangle =\sum _{l=0}^{\infty}\frac{h^{l}}{l!}\nabla _{h}^{(l)}f(h{\bf r})
:::\cdots :\int _{{\cal R}^{3}}({\bf r}^{\prime}-{\bf r})^{l}W(|{\bf r}-{\bf r}^{\prime}|,1)
d^{3}{\bf r}^{\prime},
\end{equation}
where $\nabla _{h}^{(l)}$ denotes the product of
the $\nabla$ operator with respect to coordinates $(hx,hy,hz)$ $l$ times, 
$({\bf r}^{\prime}-{\bf r})^{l}$ is a tensor of rank $l$, and the symbol
``$:::\cdots :$'' is used to denote the $l$th-order inner product. Therefore, if the
kernel approximation is to exactly reproduce a sufficiently smooth function to $(m+1)$th
order, the family of consistency relations must be fulfilled
\begin{eqnarray}
M_{0}&=&\int _{{\cal R}^{3}}W(|{\bf r}-{\bf r}^{\prime}|,1)d^{3}{\bf r}^{\prime}=1,\\
{\bf M}_{l}&=&\int _{{\cal R}^{3}}({\bf r}^{\prime}-{\bf r})^{l}W(|{\bf r}-{\bf r}^{\prime}|,1)
d^{3}{\bf r}^{\prime}={\bf 0}^{(l)},\hspace{0.5cm}{\rm for}\hspace{0.5cm}l=1,2,...,m,
\end{eqnarray}
where ${\bf 0}^{(1)}={\bf 0}=(0,0,0)$ is the null vector and ${\bf 0}^{(l)}$ is the zero
tensor of rank $l$. Fulfillment of the integral relations (5) and (6) guarantees
$C^{m}$ consistency for the kernel estimate of the function, which by virtue of 
Eq. (4) reproduces exactly the continuous function to order $m+1$. Owing to the scaling 
relation (3), the contribution of the error due to the smoothing length can be separated
from that due to the discrete representation of the integral, which, being independent
of $h$, will only depend on the number of neighbors within the kernel support and their
spatial distribution.

When solving the gravitohydrodynamics equations, gas compression is accounted for by 
evaluating the pressure gradient in the momentum equation. Therefore, consistency
relations for the kernel estimate of the gradient of a function are also of concern.
Using the definition of the kernel estimate of the gradient as
\begin{equation}
\langle\nabla f({\bf r})\rangle =\int _{{\cal R}^{3}}f({\bf r}^{\prime})
\nabla W(|{\bf r}-{\bf r}^{\prime}|,h)d^{3}{\bf r}^{\prime},
\end{equation}
expanding $f({\bf r}^{\prime})$ again in Taylor series about ${\bf r}^{\prime}={\bf r}$,
making ${\bf r}\to h{\bf r}$ and ${\bf r}^{\prime}\to h{\bf r}^{\prime}$, and inserting
the result in Eq. (7) we obtain the form
\begin{equation}
\langle\nabla_{h}f(h{\bf r})\rangle =\sum _{l=0}^{\infty}\frac{h^{l-1}}{l!}\nabla _{h}^{(l)}
f(h{\bf r}):::\cdots :\int _{{\cal R}^{3}}({\bf r}^{\prime}-{\bf r})^{l}
\nabla W(|{\bf r}-{\bf r}^{\prime}|,1)d^{3}{\bf r}^{\prime}, 
\end{equation}
where we have made use of the scaling relation 
\begin{equation}
\nabla W(h|{\bf r}-{\bf r}^{\prime}|,h)=\frac{1}{h^{\nu}}\nabla W(|{\bf r}-{\bf r}^{\prime}|,1),
\end{equation}
with $\nu =3$, which also holds for the gradient of the kernel. From Eq. (8), it follows that
$C^{m}$ consistency for the kernel estimate of the gradient is obtained only if the family
of integral relations is exactly satisfied
\begin{eqnarray}
{\bf M}^{\prime}_{0}&=&\int _{{\cal R}^{3}}\nabla W(|{\bf r}-{\bf r}^{\prime}|,1)
d^{3}{\bf r}^{\prime}=\int _{{\cal R}^{2}}W(|{\bf r}-{\bf r}^{\prime}|,1){\bf n}
d^{2}{\bf r}^{\prime}={\bf 0},\\
{\bf M}^{\prime}_{1}&=&\int _{{\cal R}^{3}}({\bf r}^{\prime}-{\bf r})
\nabla W(|{\bf r}-{\bf r}^{\prime}|,1)d^{3}{\bf r}^{\prime}={\bf I},\\
{\bf M}^{\prime}_{l}&=&\int _{{\cal R}^{3}}({\bf r}^{\prime}-{\bf r})^{l}
\nabla W(|{\bf r}-{\bf r}^{\prime}|,1)d^{3}{\bf r}^{\prime}={\bf 0}^{(l+1)},
\hspace{0.5cm}{\rm for}\hspace{0.5cm}l=2,3,...,m,
\end{eqnarray}
where ${\bf I}$ is the unit tensor. The second equality in Eq. (10) holds for any
volume enclosed by a continuous surface with differential volume element 
$d^{3}{\bf r}^{\prime}$ and differential surface element ${\bf n}d^{2}{\bf r}^{\prime}$,
where ${\bf n}$ is the outward unit normal from the volume surface. It is precisely
the requirement that the zeroth moment ${\bf M}_{0}^{\prime}={\bf 0}$, which determines
that the surface integral of the kernel must vanish identically. Because of the 
symmetry of the kernel function, relations (5) and (6) with $l$ odd are automatically
satisfied, while those with $l$ even will all appear in the expansion (4) as finite
sources of error and will not vanish unless the kernel approaches the Dirac-$\delta$
distribution. Hence, up to leading second-order Eq. (4) becomes
\begin{equation}
\langle f(h{\bf r})\rangle =f(h{\bf r})+\frac{1}{2}h^{2}\nabla _{h}\nabla _{h}
f(h{\bf r}):\int _{{\cal R}^{3}}({\bf r}^{\prime}-{\bf r})^{2}W(|{\bf r}-{\bf r}^{\prime}|,1)
d^{3}{\bf r}^{\prime}+O(h^{4}),
\end{equation}
which expresses that the kernel approximation of a function has $C^{1}$ consistency for
an unbounded domain. Using Eq. (1) it is a simple matter to show that the integral on the
right-hand side of Eq. (13), which corresponds to the second moment of the kernel
${\bf M}_{2}$, is equal to 
$\langle {\bf r}{\bf r}\rangle-\langle {\bf r}\rangle\langle {\bf r}\rangle\neq {\bf 0}^{(2)}$,
implying that $C^{2}$ consistency is not achieved even though $C^{0}$ and $C^{1}$
consistencies are automatically satisfied \citep{Sigalotti16}. This term is just the variance
of the particle position vector ${\bf r}$ and is a measure of the spread of the particle
positions relative to the mean. Evidently, 
$\langle {\bf r}{\bf r}\rangle-\langle {\bf r}\rangle\langle {\bf r}\rangle\to {\bf 0}^{(2)}$
only when $W(|{\bf r}-{\bf r}^{\prime}|,1)\to\delta ({\bf r}-{\bf r}^{\prime})$, or
equivalently, when $N\to\infty$, $h\to 0$, and $n\to\infty$.

Similarly, due to the symmetry of the kernel all integrals in Eqs. (10)-(12) will vanish
identically for $l$ even, while only those for $l$ odd will survive in the series expansion
(8), which up to second-order becomes
\begin{equation}
\langle\nabla_{h}f(h{\bf r})\rangle =\nabla _{h}f(h{\bf r})+\frac{1}{6}h^{2}\nabla _{h}
\nabla _{h}\nabla _{h}f(h{\bf r})\vdots\int _{{\cal R}^{3}}({\bf r}^{\prime}-{\bf r})^{3}
\nabla W(|{\bf r}-{\bf r}^{\prime}|,1)d^{3}{\bf r}^{\prime}+O(h^{4}),
\end{equation}
where the symbol ``$\vdots$'' is used to denote the triple inner product. Note that the
integral on the right-hand side of Eq. (14) is the third moment of the kernel gradient 
${\bf M}_{3}^{\prime}=3{\bf M}_{2}{\bf I}\neq {\bf 0}^{(4)}$. We recall that
relations (5) and (11) have important physical implications. In particular, satisfaction
of relation (5) means that the homogeneity of space is not affected by the SPH kernel
approximation, which has as a consequence the conservation of linear momentum. On the
other hand, fulfillment of relation (11) expresses that the isotropy of space is
preserved by the kernel approximation, and therefore angular momentum is locally conserved 
\citep{Vignjevic09,Sigalotti16}. 

An important feature of Eqs. (13) and (14) is that the contribution of $h$ to the error
can be separated from the error carried by the discretization of the consistency relations,
which will only depend on the number of neighbors, $n$, and how they are distributed within
the kernel support. In general, it is well-known that the particle approximation of Eq. (5)
diverges from being exactly one, i.e.,
\begin{equation}
M_{0,a}=\sum _{b=1}^{n}W_{ab}\Delta V_{b}\neq 1,
\end{equation}
where $W_{ab}=W(|{\bf r}_{a}-{\bf r}_{b}|,h)$ and $\Delta V_{b}$ is the volume of the subdomain
of neighbor particle $b$. The error carried by Eq. (15) scales as $\sim n^{-\gamma}$,
with $\gamma\in [0.5,1]$, depending on the particle distribution \citep{Monaghan92,Zhu15}.
Therefore, as the number of neighbors is increased the discrete normalization condition
approaches unity and $C^{0}$ particle consistency is restored. This will also make the
particle approximations of relations (10) and (11) to approach the null vector ${\bf 0}$
and the unit tensor, respectively. These  conditions state that the particles should
provide a good approximation to a partition of unity. As in traditional
finite difference and finite element methods, the concept of consistency in SPH defines how
well the discrete model equations represent the exact equations in the continuum limit. In
SPH this is accomplished in two separate steps: the kernel approximation which, as we have
described above, is derived from the continuous form, and the particle approximation, where the
integrals are replaced by sums over a finite set of particles within the kernel support. Since
the kernel consistency relations do not assure consistency for the particle approximation,
the discrete counterparts of Eqs. (13) and (14) must be written as
\begin{eqnarray}
f_{a}\to\langle f\rangle _{a}&=&M_{0,a}(f)_{a}+h(\nabla f)_{a}\cdot {\bf M}_{1,a}+\frac{1}{2}h^{2}
(\nabla\nabla f)_{a}:{\bf M}_{2,a}+O(h^{3}),\\
\nabla _{a}f_{a}\to\langle\nabla f\rangle _{a}&=&\frac{1}{h}{\bf M}_{0,a}^{\prime}(f)_{a}
+(\nabla f)_{a}\cdot
{\bf M}_{1,a}^{\prime}+\frac{1}{2}h(\nabla\nabla f)_{a}:{\bf M}_{2,a}^{\prime}+
\frac{1}{6}h^{2}(\nabla\nabla\nabla f)_{a}:\cdot {\bf M}_{3,a}^{\prime}+
O(h^{3}),
\end{eqnarray}
respectively, where quantities between parentheses denote exact values of the function and 
its derivatives at the position of particle $a$ and the particle representation of the
consistency integrals is given by
\begin{eqnarray}
M_{0,a}&=&\sum _{b=1}^{n}W_{ab}\Delta V_{b},\\
{\bf M}_{l,a}&=&\sum _{b=1}^{n}{\bf r}_{ba}^{l}W_{ab}\Delta V_{b},\hspace{0.5cm}{\rm for}
\hspace{0.5cm}l=1,2,\\
{\bf M}_{0,a}^{\prime}&=&\sum _{b=1}^{n}\nabla _{a}W_{ab}\Delta V_{b},\\
{\bf M}_{l,a}^{\prime}&=&\sum _{b=1}^{n}{\bf r}_{ba}^{l}\nabla _{a}W_{ab}\Delta V_{b},
\hspace{0.5cm}{\rm for}\hspace{0.5cm}l=1,2,3,
\end{eqnarray}
where ${\bf r}_{ba}={\bf r}_{b}-{\bf r}_{a}$ and $\Delta V_{b}=m_{b}/\rho _{b}$, with
$m_{b}$ and $\rho _{b}$ denoting the mass and density of particle $b$, respectively.
According to Eqs. (16) and (17), $C^{0}$ particle consistency for the function and its
gradient will demand that $M_{0,a}=1$, ${\bf M}_{0,a}^{\prime}={\bf 0}$, and
${\bf M}_{1,a}^{\prime}={\bf I}$ at the position of particle $a$, while $C^{1}$
particle consistency is restored if in addition ${\bf M}_{1,a}={\bf 0}$ and
${\bf M}_{2,a}^{\prime}={\bf 0}^{(3)}$ are satisfied. In particular, restoring $C^{0}$
particle consistency implies that the homogeneity and isotropy of the discrete space
is preserved, which has as a consequence the conservation of linear and angular 
momentum in practical calculations \citep{Vignjevic09,Sigalotti16}. The goal here is to 
track the quality of the
particle consistency relations in a true hydrodynamic evolution involving large
density and pressure gradients as well as large spatial and temporal variations of
the smoothing length. This will allow us to evaluate the degree of consistency that can
be achieved when the number of neighbors within the kernel support and the smoothing
length are allowed to vary with $N$ according to the scalings $n\sim N^{1/2}$ and
$h\sim N^{-1/6}$, which approach asymptotically the joint limit $N\to\infty$,
$h\to 0$, and $n\to\infty$ for particle consistency as $N$ is increased.

\section{SPH solver} \label{sec:solver}

A modified version of the simulation code GADGET-2 is used for the calculations of this
paper. The code relies on a fully conservative formulation where the discrete Euler
equations are derived via a variational principle from the discretized Lagrangian of
the fluid system \citep{Springel02,Springel05}. As in most SPH formulations, the density
estimate is calculated by the summation interpolant
\begin{equation}
\rho _{a}=\sum _{b=1}^{n}m_{b}W_{ab},
\end{equation}
while the Euler-Lagrange equations of motion for the particles are given by
\begin{equation}
\left(\frac{d{\bf v}_{a}}{dt}\right)_{SPH}=-\sum _{b=1}^{n}m_{b}\left[\frac{f_{a}p_{a}}
{\rho _{a}^{2}}\nabla _{a}W_{ab}(h_{a})+\frac{f_{b}p_{b}}{\rho _{b}^{2}}
\nabla _{b}W_{ab}(h_{b})\right],
\end{equation}
where ${\bf v}_{a}$ and $p_{a}$ are the particle velocity and pressure, respectively,
and the factor $f_{a}$ is defined by
\begin{equation}
f_{a}=\left(1+\frac{h_{a}}{3\rho _{a}}\frac{\partial\rho _{a}}{\partial h_{a}}
\right)^{-1}.
\end{equation}
The velocity of particle $a$ is then updated according to
\begin{equation}
\frac{d{\bf v}_{a}}{dt}=\left(\frac{d{\bf v}_{a}}{dt}\right)_{SPH}+
\left(\frac{d{\bf v}_{a}}{dt}\right)_{GRAV}+\left(\frac{d{\bf v}_{a}}{dt}\right)_{AV},
\end{equation}
where the last two terms on the right-hand side account for the self-gravitational
acceleration and the artificial viscous forces, respectively.
The gravitational forces are calculated using a hierarchical multipole expansion,
which can be applied in the form of a TreePM method, where short-range forces are
calculated with the tree method and long-range forces are determined using mesh-based
Fourier methods. A detailed account of the code is given by
\cite{Springel05}. Here we shall only briefly describe the improvements that have
been incorporated in our version of the code.

One straightforward way of restoring particle consistency and therefore reducing the
zeroth-order error terms carried by the SPH representation of the continuity and
momentum equations \citep{Read10} is just to increase the number of
particles within the kernel support. However, conventional kernels, like the widely
used cubic $B$-spline kernel of \cite{Monaghan85}, suffer from a pairing instability
when working with large numbers of neighbors, where particles come into close pairs
and become less sensitive to small perturbations within the kernel support 
\citep{Dehnen12,Price12,Hayward14}. To overcome this difficulty, we have adopted
a Wendland C$^{4}$ kernel function \citep{Wendland95,Dehnen12} 
\begin{equation}
W(q,h)=\frac{495}{32\pi h^{3}}\left(1-q\right)^{6}\left(1+6q+\frac{35}{3}q^{2}\right),
\end{equation}
if $q\leq 1$ and $0$ otherwise, where $q=|{\bf r}-{\bf r}^{\prime}|/h$. As was demonstrated
by \cite{Dehnen12}, Wendland functions have positive Fourier transforms and so they can
support arbitrarily large numbers of neighbors without favoring a close pairing of
particles. Moreover, the exact particle distribution depends on the dynamics of the
flow and on the kernel function that is employed. This makes the accuracy assessment
of SPH a non-trivial problem. However, Wendland functions are very reluctant to allow
for particle motion on a sub-resolution scale and, in contrast to most commonly used
kernels, they maintain a very regular particle distribution, even in highly dynamical
tests \citep{Rosswog15}. 

A further improvement includes the update of the artificial viscosity switch using 
the method proposed by \cite{Hu14}. In this method the artificial viscosity term
entering on the right-hand side of Eq. (25) is implemented as in GADGET-2 by the
common form \citep{Monaghan97}
\begin{equation}
\left(\frac{d{\bf v}_{a}}{dt}\right)_{AV}=-\sum _{b=1}^{n}m_{b}\Pi _{ab}\nabla _{a}\bar W_{ab},
\end{equation}
where $\bar W_{ab}=[W_{ab}(h_{a})+W_{ab}(h_{b})]/2$ and
\begin{equation}
\Pi _{ab}=-\frac{1}{2}\frac{\bar\alpha _{ab}v_{\rm sig}}{\bar\rho _{ab}}\omega _{ab}
\hspace{0.5cm}{\rm if}\hspace{0.5cm}\omega _{ab}<0,
\end{equation}
and zero otherwise. Here 
$\omega _{ab}=({\bf v}_{a}-{\bf v}_{b})\cdot {\bf r}_{ab}/|{\bf r}_{ab}|$,
$v_{\rm sig}=c_{a}+c_{b}-3\omega _{ab}$ is the signal speed, $c_{a}$ is the particle
sound speed, $\bar\rho _{ab}=(\rho _{a}+\rho _{b})/2$, and 
$\bar\alpha _{ab}=(\alpha _{a}+\alpha _{b})/2$. We note that in the original GADGET-2
code formulation $\alpha _{a}=\alpha _{b}=const$. It is well-known that this form of
the artificial viscosity introduces excessive dissipation in shear flows, leading to
spurious angular momentum transport in the presence of vorticity. Therefore, it is
desirable to suppress this excessive dissipation in regions where the vorticity 
dominates over the velocity divergence \citep{Nelson09}. In particular, \cite{Morris97}
proposed individual viscosity coefficients that adapt their values according to
velocity-based source terms. Later on, \cite{Cullen10} improved on this formulation
by devising a novel shock indicator based on the total time derivative of the velocity
divergence, which distinguishes shocks from purely convergent flows and discriminates 
between pre- and post-shocked regions. While this prevents false triggering of the
artificial viscosity, their method includes a limiter which puts a stronger weight
on the velocity divergence than on the vorticity. The artificial viscosity switch
proposed by \cite{Hu14} follows the same principles of that presented by \cite{Cullen10},
except that it now uses a limiter that applies the same weight to the velocity divergence
and vorticity. The method consists of calculating the viscosity coefficient through
the following steps. A target value of the viscosity coefficient is first calculated
using the relation
\begin{equation}
\alpha _{{\rm tar},a}=\alpha _{\rm max}\frac{h_{a}^{2}S_{a}}{h_{a}^{2}S_{a}+c_{a}^{2}},
\end{equation}
where $\alpha _{\rm max}=0.75$ and $S_{a}=\max (0,-\dot {\nabla}\cdot {\bf v}_{a})$ is 
the shock indicator. The total time derivative of the velocity divergence is given by
$-\dot {\nabla}\cdot {\bf v}=d^{2}\ln\rho /dt^{2}$ after differentiation of the continuity
equation and the divergence of the velocity is evaluated using the higher-order estimator
proposed by \cite{Cullen10}. Hence $\dot {\nabla}\cdot {\bf v}<0$ is indicative of nonlinear
flow steepening as occur in pre-shocked regions, while in post-shocked regions 
$\dot {\nabla}\cdot {\bf v}>0$. The true viscosity coefficient that enters in Eq. (28) is
then defined by
\begin{equation}
\alpha _{a}=\left\{
\begin{array}{ll}
\xi _{a}\alpha _{{\rm tar},a}\hspace{5.1cm}{\rm if}\hspace{0.5cm}\alpha _{a}\leq\alpha _{{\rm tar},a},\\
\xi _{a}\left[\alpha _{{\rm tar},a}+\left(\alpha _{a}-\alpha _{{\rm tar},a}\right)
\exp\left(-\frac{\Delta t}{\tau _{a}}\right)\right]\hspace{0.5cm}{\rm if}\hspace{0.5cm}
\alpha _{a}>\alpha _{{\rm tar},a},
\end{array}
\right.
\end{equation}
where $\xi _{a}$ is a modified limiter given by
\begin{equation}
\xi _{a}=\frac{|(\nabla\cdot {\bf v})_{a}|^{2}}{|(\nabla\cdot {\bf v})_{a}|^{2}+
|(\nabla\times {\bf v})_{a}|^{2}+0.0001(c_{a}/h_{a})^{2}},
\end{equation}
$\Delta t$ is the time step, and $\tau _{a}=10h_{a}/v_{\rm sig}$ is the decay time with a decay 
speed equal to
\begin{equation}
v_{\rm decay}=\max _{|{\bf r}_{ab}|\leq h_{a}}[\bar c_{ab}-\min (0,\omega _{ab})],
\end{equation}
where $\bar c_{ab}=(c_{a}+c_{b})/2$. This method, which is referred to as an artificial
viscosity with a strong limiter, suppresses viscous dissipation in subsonically 
convergent flows and ensures that $\alpha _{a}$ rises rapidly up to $\alpha _{\rm max}$
when the converging flow becomes supersonic. This is a desirable property in protostellar
collapse simulations where holding $\alpha _{a}$ to a fixed constant value during the
evolution may cause unphysical dissipation of local velocity differences away from shocks.
Such adverse effects of the artificial viscosity are responsible for the oversmoothing of 
weak shocks as well as the damping of adiabatic oscillations and shear flows, thereby
seriously affecting the outcome of the simulations.

\section{Test problems} \label{sec:tests}

\subsection{Particle consistency relations for a set of points}

We first test the quality of the first few moments in relations (18)-(21) for a static
set of $N=64^{3}$ points distributed within a cube of length $L=1$, density $\rho =1$,
pressure $p=1$, and sound speed $c^{2}=\gamma p/\rho$, with $\gamma =5/3$. Two different 
particle distributions are considered: a glass-like
configuration and a random distribution. A similar test problem was employed by
\cite{Zhu15} to determine the quality of the discrete normalization condition given by
relation (18) as the number of neighbors is increased from $n=48$ to 3200.
Here the glass distribution was obtained from GADGET-2 by enabling the corresponding
code option and starting with {\bf a random distribution} of SPH particles
and an expansion factor $a=0.01$ \citep{Couchman95,White96}. The resulting outcome is
then evolved hydrodynamically up to a time $t=1.1$ in code units, using an isothermal 
equation of state, periodic boundary conditions at the edges of the box, and excluding
self-gravity. As the system evolves toward a relaxed state, the distributions of quantities,
such as the smoothing length, the density, and the discrete moments given by relations
(18)-(21) tend to normal distributions. The evolution ends up with an equilibrium
configuration in which all particles have approximately equal SPH densities ($\approx 1$).

{\bf In real SPH applications, the distances between neighboring particle pairs tend
to equilibrate due to pressure forces, which makes the interpolation errors much smaller
and the irregularity of the particle distribution more ordered than for a random
distribution, where particles sample the fluid in a
Poissonian fashion. In this sense, a random configuration represents an extreme case
for SPH simulations. In contrast, a glass configuration mimics the other extreme case
where the particle distribution is quasi-regular and almost force-free. Although a
random distribution rarely occurs in SPH, except perhaps in highly turbulent flows where
particles are highly disordered and SPH is unable to re-order the particles, we analyze
the quality of the density estimate and discrete moments of the kernel (and kernel
gradient) for a random distribution with the only purpose of comparing with the results
obtained by \cite{Zhu15}}.

The top and bottom panels of Figure 1 show histograms of the particle density estimate
[Eq. (22)] with increasing $n$ from 48 to 3200 as in \cite{Zhu15} for the glass and
random particle distributions, respectively. As expected, the density distribution for
the glass configuration is much narrower than for the randomly distributed points. In
the former case, as $n$ is increased the density distribution approaches a Dirac-$\delta$
distribution, while in the latter case, the distribution slowly approaches a Gaussian
shape with a peak at $\rho =1$. Also, for small values of $n$ the density distribution
shows long tails for the random configuration which are not present for the glass distribution,
at least when $n\geq 120$. \cite{Zhu15} argued that such long tails are due to an overestimate
of the density produced by the particle self-contribution in Eq. (22) when the SPH particles
are randomly spaced. {\bf While this result is well-known, it can also be derived analytically
from Eq. (22) given the mass of the particles, their number and actual distribution within 
the kernel volume, and the form of the kernel function}. {\bf However, using the M$_{4}$
kernel of \cite{Monaghan85}, \cite{Whitworth95} demonstrated that the overestimate in
density occurs because a random distribution produces a fluctuating density field, where
the particle positions are correlated with the overdense fluctuations and anticorrelated
with the underdense fluctuations. In other words, the expectation value of the density at
the location of a particle will be overestimated by a value almost exactly equal to the
``self-density''. This is the reason why it seems appropriate to exclude the particle
self-contribution from Eq. (22). \cite{Whitworth95} concluded that as soon as the particle
positions are settled before they are allowed to evolve dynamically, exclusion of the
self-contribution will lead to a significant error because the particle distribution will
cease to be random and the density fluctuations will be removed}.
Table 1 lists the standard deviation, $\sigma (\rho)$, and expectation 
value, $\langle\rho\rangle _{e}$, of the density measured from the distributions of Figure 1.
For both particle configurations, $\sigma (\rho)$ decreases with increasing $n$, while
the value of $\langle\rho\rangle _{e}$ becomes close to unity when $n\gtrsim 480$
for a glass configuration and $n\gtrsim 1600$ for a random distribution. This agrees with
the $\sim n^{-1/2}$ and $n^{-1}$ trends of $\sigma (\rho)$ as a function of $n$ found by 
\cite{Zhu15} for a truly random and a glass-like configuration, respectively. Since
the results of Figure 1 are consistent with the findings of \cite{Zhu15} for these
tests, we feel confident to proceed with a similar statistical analysis to measure the
quality of the discrete moments as $n$ is increased.

According to the series expansions (16) and (17), the error in the density and density
gradient estimates separates into two contributions: one due to the local value of $h$
and the other due to the discrete values of the moments.
Since the latter are independent of $h$, they will only depend on the number of
neighbors within the kernel support. This observation introduces a subtle difference 
between the meaning of consistency and accuracy in SPH. Consistency demands that
$n\to\infty$, while accuracy demands that $n\to\infty$ and $h\to 0$ in order to have
convergent results in the limit $N\to\infty$. Therefore, we can achieve approximate particle 
consistency and improved accuracy as $n$ is increased and $h$ is decreased with $N$. 

To achieve $C^{0}$ consistency, the parameter $M_{0,a}$ and the mean of the elements of
matrix ${\bf M}_{1,a}^{\prime}$ should peak around 1, while the mean of
the components of vector ${\bf M}_{0,a}^{\prime}$ should follow a peaked distribution
around 0. Moreover, $C^{1}$ consistency will additionally require that the mean of the
components of ${\bf M}_{1,a}$ and the mean of
the elements of matrix ${\bf M}_{2,a}^{\prime}$ both peak around 0. The distributions of all
these quantities are plotted in Figures 2 and 3 for the glass and random configurations,
respectively. We may see that for the glass configuration the distributions follow the
desired behavior and approach a Dirac-$\delta$ function as $n$ is increased, indicating
that approximate $C^{1}$ consistency is achieved for the density and its gradient when
$n=3200$. Conversely, for a random configuration the distributions approach Gaussian-like
shapes with peaks close to the continuum values. For both particle configurations
the distributions of $M_{0,a}$ peak at values lower than 1 for small $n$, suggesting an
overestimation of the particle density for small numbers of neighbors. Table 1 lists the
standard deviations and expectation values of these moments as calculated by fitting a Gaussian
function to the histograms of Figures 2 and 3. From these values we see that very good
$C^{0}$ and $C^{1}$ consistencies are achieved for the glass configuration. A stronger
sensitivity to the particle distribution is observed for ${\bf M}_{0,a}^{\prime}$, which
is exacerbated for the random configuration. In this case, the standard deviation of the
distribution is consistently larger for smaller $n$ and converges to zero
rather slowly compared to the glass configuration. Evidently, ${\bf M}_{0,a}^{\prime}$
seems to be more sensitive to the degree of particle disorder than the other parameters for
this test, implying a higher error in the SPH representation of the gradient. Although this error
depends on the quality of the particle distribution, it can be regulated by further
increasing the number of neighbors since the standard deviation is expected to follow a
trend between $n^{-1/2}$ and $n^{-1}$ in actual SPH calculations. Also, note that the
expectation values of ${\bf M}_{1,a}$, ${\bf M}_{0,a}^{\prime}$, and ${\bf M}_{2,a}^{\prime}$
are always zero because of the symmetry of the kernel.

The required computational cost in CPU time for a complete run is nearly the same
for the standard and modified GADGET-2 code. However, the computational cost increases
almost linearly with $n$ for fixed $N$. For instance, a run with $n=3200$ took about
11 s compared to $\sim 0.34$ s for $n=64$, implying a factor of $\sim 32$ more CPU
time. Thus, increasing the number of neighbors increases the computational cost, which
is the price that has to be paid when SPH is used as a numerically consistent method.
Also, \cite{Zhu15} argued that when $n$ is made to vary with $N$ as $N^{0.5}$, the
computational cost scales with $N$ as $O(N^{1.5})$ rather than as $O(N)$ as for
traditional choices of $n$. 

\subsection{Two-dimensional Keplerian ring}

We now test the performance of our implemented artificial viscosity for an equilibrium 
ring of isothermal gas rotating about a central point mass. This test is the same documented
by \cite{Cullen10} and \cite{Hu14}. Self-gravity of the ring is neglected and perfect
balance between pressure forces, gravitational attraction from the central point mass,
and centrifugal forces is assumed. The surface density of the gas is given by the Gaussian
profile
\begin{equation}
\Sigma (r)=\frac{1}{m}\exp\left[\frac{-(r-r_{0})^{2}}{2\sigma ^{2}}\right],
\end{equation}
where $m$ is the mass of the ring, $r$ is the radial distance from the central point
mass ($r=0$), $r_{0}=10$, and $\sigma =1.25$ is the width of the ring. For the central
point mass we set $GM=1000$, where $G$ is the gravitational constant and $M\gg m$. The
ring is filled with $N=9987$ particles, initially distributed using the method of
\cite{Cartwright09}. With these parameters, the ring is in differential rotation with
an azimuthal velocity $v_{\phi}=\sqrt{GM/r}=10$ and a rotation period $T=2\pi$ at 
$r=r_{0}$. The sound speed is set to $c=0.01$. This value is much smaller than the
azimuthal velocity so that dynamical instabilities in the ring are expected to occur 
only after many rotation periods \citep{Cullen10}. Under Keplerian differential
rotation the flow is shearing and therefore any viscosity may cause the ring to break
up \citep{Lynden74}, with the instability initiating at its inner edge
\citep{Maddison96}.

Figure 4 shows the ring configuration as obtained using four different SPH schemes.
The times are given in code units. When GADGET-2 is used with the cubic 
$B$-spline kernel and $n=12$ neighbors together with the standard artificial
viscosity formulation with a constant coefficient $\alpha =0.8$ (Fig. 4a), the ring becomes
unstable at $t\approx 3.8$ (corresponding to $\approx 0.6T$). At $t=12$, i.e., after
approximately two rotation periods, the inner edge instability is well-developed and the
ring is close to break up. When the same run is repeated using \cite{Hu14} scheme for the
artificial viscosity with the higher-order velocity divergence estimator proposed by
\cite{Cullen10} and $\alpha$ varying in the interval $[0,0.8]$, the instability manifests
in the form of particle clusterings and voids in the particle distribution,
resembling a sort of sticking instability, as shown in Figure 4b at $t=49$ (corresponding
to $\approx 7.8$ rotation periods). The calculation stops soon thereafter because
of failure of the TreePM algorithm when clustered particles become too close to 
one another. Only
little improvement is obtained when using the standard artificial viscosity formulation and 
the Wendland C$^{4}$ function with $n=120$ neighbors as the ring becomes unstable after
about one rotation period ($t\approx 6.6$). Figure 4c shows progress of the instability
at a later time ($t=15$). Therefore, changing the cubic $B$-spline kernel with a
Wendland C$^{4}$ function with $n=120$, or even larger $n$, while maintaining the standard
artificial viscosity causes the instability to grow a little more slowly. Only when
this latter run is repeated using \cite{Hu14} scheme for the artificial viscosity does
the ring stay stable for more than 20 rotation periods (Fig. 4d). The ring preserves its
particle configuration and remains stable even when the evolution is followed for more
than 30 rotation periods.

\section{Protostellar collapse simulations} \label{sec: collapse}

We now test the consistency and accuracy of our implemented SPH method for
numerical hydrodynamical calculations involving large density and pressure gradients as well 
as variable smoothing lengths. As a problem we choose the collapse and fragmentation of an
isolated molecular cloud core. The templates for the model clouds correspond to the
well-known standard isothermal test case in the variant calculated by \cite{Burkert93} and
the centrally condensed, Gaussian cloud advanced by \cite{Boss91}.

\subsection{Initial conditions}

\subsubsection{Standard isothermal cloud}

The standard isothermal test case starts from a uniform density ($\rho _{0}=3.82\times 10^{-18}$
g cm$^{-3}$) sphere of mass $M=1M_{\odot}$, radius $R=4.99\times 10^{16}$ cm, temperature
$T=10$ K, and solid-body rotation $\omega =7.2\times 10^{-13}$ s$^{-1}$. The model has
ideal gas thermodynamics with a mean molecular weight $\mu\approx 3$ and an isothermal
sound speed $c_{\rm iso}\approx 1.66\times 10^{4}$ cm s$^{-1}$. With these parameters the
initial mean free-fall time is $t_{\rm ff}\approx 1.07\times 10^{12}$ s. In order to favor 
fragmentation into a binary system, the uniform density background is perturbed azimuthally
as
\begin{equation}
\rho =\rho _{0}\left[1+0.1\cos(2\phi)\right],
\end{equation}
where $\phi$ is the angle about the spinning $z$-axis. With these parameters the ratios of
thermal and rotational energies to the absolute value of the gravitational energy are
$\alpha\approx 0.26$ and $\beta\approx 0.16$, respectively.

\subsubsection{Gaussian cloud}

The Gaussian cloud corresponds to a centrally condensed sphere of the same mass and radius as
the standard isothermal cloud. The radial central condensation is given by
\begin{equation}
\rho (r)=\rho _{\rm c}\exp\left[-\left(\frac{r}{b}\right)^{2}\right],
\end{equation}
where $\rho _{\rm c}=1.7\times 10^{-17}$ g cm$^{-3}$ is the initial central density and
$b\approx 0.578R$. This produces a central density 20 times higher than the density at the
outer edge. Solid-body rotation is assumed at the rate $\omega =1.0\times 10^{-12}$ s$^{-1}$.
The gas has a temperature of 10 K, a chemical composition corresponding to a mean molecular
weight $\mu\approx 2.28$, and an isothermal sound speed $c_{\rm iso}\approx 1.90\times 10^{4}$
cm s$^{-1}$. The central free-fall time is $t_{\rm ff}\approx 5.10\times 10^{11}$ s and the
radial density distribution is azimuthally perturbed using Eq. (34). With this choice of the
parameters, the values of $\alpha$ and $\beta$ are the same as for the uniform-density,
standard isothermal test.

\subsubsection{Equation of state}

A barotropic pressure-density relation of the form \citep{Boss00}
\begin{equation}
p=c_{\rm iso}^{2}\rho +K\rho ^{\gamma},
\end{equation}
is used for both the uniform- and Gaussian-cloud models, where $\gamma =5/3$ and $K$ is a
constant determined from equalizing the isothermal and adiabatic parts of Eq. (36) at a 
critical density $\rho _{\rm crit}=5.0\times 10^{-12}$ g cm$^{-3}$ for the isothermal test
case and $5.0\times 10^{-14}$ g cm$^{-3}$ for the Gaussian cloud, which separates the
isothermal from the nonisothermal collapse. The local sound speed is therefore given by
\begin{equation}
c^{2}=c_{\rm iso}^{2}\left[1+\left(\frac{\rho}{\rho _{\rm crit}}\right)^{\gamma -1}\right],
\end{equation}
so that $c\approx c_{\rm iso}$ when $\rho\ll\rho _{\rm crit}$ and 
$c\approx\gamma ^{1/2}c_{\rm iso}$ when $\rho\gg\rho _{\rm crit}$. With these choices of
the critical density we allow the standard isothermal cloud to evolve deep into the
isothermal collapse to provide direct comparison with previous barotropic SPH calculations
by \cite{Kitsionas02} and \cite{Arreaga07}, while a value of 
$\rho _{\rm crit}=5.0\times 10^{-14}$ g cm$^{-3}$ produces a behavior that is more
representative of the near-isothermal phase and fits better the Eddington approximation
solution of \cite{Boss00}.

\subsubsection{Initial particle distribution and smoothing length}

All collapse calculations start from a set of points in a glass configuration, which was
generated from {\bf randomly distributed particles} using the GADGET-2 glass-making mode. As shown
in Table 2, we consider two separate sequences of calculations with varying total number
of particles ($N$) for both the uniform and Gaussian cloud models. Models labeled U1C-U4C
correspond to uniform clouds calculated with the standard GADGET-2 using a fixed number of
neighbors ($n=64$), while models U1W-U4W were calculated using our modified GADGET-2 code
using a Wendland C$^{4}$ function with varying number of neighbors. Similarly, models
G1C-G6C and G1W-G6W correspond to Gaussian clouds using the standard (with $n=64$) and
modified code (with varied $n$), respectively.

For these tests, we use the parameterization provided by \cite{Zhu15}, where $h$ is
allowed to vary with $N$ as $h\propto N^{-1/6}$. With this choice we obtain the scaling
relations $n\approx 7.61N^{0.503}$ and  $h\approx 7.23n^{-0.33}$ so that $h$ decreases as
the number of neighbors increases. Thus, choosing the proportionality factor of the scaling
$h\propto N^{-1/6}$ as exactly unity gives an exponent for the dependence of $h$ on
$n$ that is close to the suggested value of $-1/3$. The variation of $h$ with $n$
is depicted in Figure 5. For small values of $n$ the smoothing length decreases
rapidly as $n$ increases and then more slowly at larger values of $n$, asymptotically
approaching zero as $n\to\infty$ as required to restore particle consistency.

We note that models U1W--U4W do not satisfy the Jeans condition for densities above
$\approx 5.0\times 10^{-14}$ g cm$^{-3}$ due to their much larger numbers of neighbors
compared to models U1C--U4C. In contrast, models G1W--G6W all meet the Jeans resolution
requirements for gravitational fragmentation. However, in order to avoid spurious
fragmentation the gravity softening length of each particle, $\epsilon _{a}$, is
evolved with time in step with its corresponding smoothing length $h_{a}$ so that
$\epsilon _{a}\approx h_{a}$ \citep{Bate97}. In addition, \cite{Hubber06} showed
that SPH reproduces the analytical Jeans criterion and simulates gravitational
fragmentation properly, even at very poor resolution. That is, artificial fragmentation
is suppressed in regions where the Jeans mass is less than the minimum resolvable mass,
$M_{\rm min}=nm$, provided the standard kernel-softened gravity ($\epsilon\approx h$)
is used, where $m$ is the mass of a single SPH particle. This way unresolved 
Jeans-unstable condensations are stabilized numerically. Thus, \cite{Hubber06} concluded
that failing to satisfy the Jeans condition simply suppresses true fragmentation in
SPH calculations, rather than resulting in artificial fragmentation as in
finite-difference codes. Similar conclusions were previously met by \cite{Whitworth98}
through an analytical derivation of the Jeans criterion.

\subsection{Collapse of the uniform cloud}

Although a uniform-density profile is an extreme idealization of a real cloud core,
it provides a simple model to learn how nonaxisymmetric perturbations grow from a
structureless medium. Perhaps the most illustrative example of this is given by the
standard isothermal test case, which was originally proposed by \cite{Boss79} and
thereafter used as a benchmark for testing numerical codes studying protostellar
collapse and fragmentation processes, with the fairly good agreement that the outcome
of the first evolution is the formation of a protostellar binary system
\citep{Burkert93,Truelove98,Boss00,Kitsionas02,Springel05,Arreaga07}. Previous 
highly-resolved SPH calculations for this test over $\sim 9$ orders of magnitude increase
in density and using a limited number of neighbors ($n\approx 64$) have predicted the
formation of two elongated fragments connected by a filamentary bar when the maximum
density in the fragments has passed $\rho _{\rm crit}=5.0\times 10^{-12}$ g cm$^{-3}$
\citep{Kitsionas02,Arreaga07}. When the gas within the fragments becomes adiabatic
and heats up, their cylindrical collapse slows down. This makes the fragments to
approach a rather spherical shape, while the connecting bar, which remains isothermal,
collapses to a singular filament with no signs of fragmentation. However,
comparisons between all these earlier calculations have been performed with varied 
total numbers of particles $N$ and a constant number of neighbors $n\approx 64$ or so,
and therefore they are likely to suffer from a loss of consistency due to persisting
zeroth-order discretization errors, whose magnitudes may even grow at a faster rate 
when approaching the limit $N\to\infty$ and $h\to 0$ \citep{Read10}.

Figure 6 displays column density images of the cloud midplane during the collapse
of model U4C using the original GADGET-2 formulation with $n=64$ neighbors. We may
see that up to $1.2736t_{\rm ff}$ (peak density of $\sim 10^{8.91}\rho _{0}$), the
morphology of collapse and the fragmentation details are very similar to previously
reported SPH results for this model. A singular bar connecting two quasi-spherical
fragments is formed and the details of the fanning-out of the bar close to the binary
fragments are also reproduced. However, when the calculation is continued farther in time 
the binary components undergo rapid rotational disruption into smaller fragments (see
the last snapshot at $1.302t_{\rm ff}$ when the peak density is $\sim 10^{9.74}\rho _{0}$).
Meanwhile the gas within the singular bar becomes adiabatic, hindering its cylindrical
collapse and fragmenting along its length into similar small objects. Due to their excess kinetic
energy acquired during rotational disruption of the former binary components, some of these
fragments collide and merge between them and/or with those coming from the bar breakup,
followed by a rather chaotic dynamics at later times. At these stages, the outcomes of
models U1C--U4C show no sign of convergence at comparable maximum densities. {\bf However,
we note that the lack of convergence is not surprising because at this stage the small-scale
fragmentation observed derives from the non-linear amplification of particle noise inherent
in SPH, which leads to different patterns as the spatial resolution is increased. This noise
arises because mutually repulsive pressure forces between particle pairs do not cancel
in all directions simultaneously. It affects the accuracy of SPH and leads to slow
convergence rates. On the other hand, the use of the standard artifical viscosity with a
constant coefficient leads to spurious angular momentum transport in the presence of
vorticity, which may cause the rotational disruption of the binary fragments}.

The time evolution of the distribution of the first few moments given by
relations (18)--(21) is depicted in Figure 7 for model U4C. Only the late stages
of collapse during the process of fragmentation are shown. All plots represent only
particles carrying a density greater than $\rho _{\rm crit}$ as identified from the
last snapshot generated during the simulation. Starting from a given point in the
evolution, {\bf histograms} for the particle density, smoothing length, and moment distributions
are constructed. The gray strips in the plots of Figure 7 correspond to the time evolution 
of the standard deviations calculated with respect to the maximum of the distributions
(marked with the solid lines), where most particles lie. Hence, the width of the
strips at any given time gives the width of the corresponding distribution. This procedure
allows us to evaluate the quality of the consistency relations in rapidly evolving
regions of high density where the smoothing length is also varying rapidly to guarantee
adequate spatial resolution. According to expansions (16) and (17), the trends of the 
$M_{0}$, $\langle {\bf M}_{0}^{\prime}\rangle$, and $\langle {\bf M}_{1}^{\prime}\rangle$ 
distributions are indicative of whether $C^{0}$ consistency is achieved during the
evolution, while the degree of $C^{1}$ consistency is measured by the time evolution of the 
$\langle {\bf M}_{1}\rangle$ and $\langle {\bf M}_{2}^{\prime}\rangle$ distributions.
The maxima of the distributions of $\langle {\bf M}_{1}\rangle$ and
$\langle {\bf M}_{2}^{\prime}\rangle$ always peak at zero because of the symmetry of
the kernel. Note that the maxima of the $\langle {\bf M}_{0}^{\prime}\rangle$ distribution
also peak at zero, except toward the end of the evolution when they start to oscillate
erratically about a mean value close to zero. As the smoothing length decreases sharply,
within the growing fragments, the width of $\langle {\bf M}_{1}\rangle$ and
$\langle {\bf M}_{2}^{\prime}\rangle$ contracts until approaching a Dirac-$\delta$
distribution. In contrast, the width of $\langle {\bf M}_{0}^{\prime}\rangle$
remains approximately constant. On the other hand, the peaks of the distributions of
$M_{0}$ and $\langle {\bf M}_{1}^{\prime}\rangle$ are always below unity, meaning that
$C^{0}$ consistency is not achieved. The deviations from unity of $M_{0}$ are even larger
than those of $\langle {\bf M}_{1}^{\prime}\rangle$, implying that the estimate of the
function is more sensitive to the particle discretization errors than the estimate of 
the gradient. Violation of the normalization condition by the particle
approximation means that the calculation of model U4C is even {\bf worse} than first-order
accurate. Similar temporal variations of the estimates of the moments were also observed for
models U1C--U3C at lower resolution. For all these models, the values of $M_{0}$ and
$\langle {\bf M}_{1}^{\prime}\rangle$, which were initially closer to unity, degraded 
gradually in the course of collapse and the time interval represented in Figure 7
corresponds to that of maximum deviation. 

Details of the evolution of models U2W and U4W
are shown in Figures 8 and 9, respectively, at comparable maximum densities.
As shown in Table 2, model U2W is run with $N=600000$ and $n=6121$, while model
U4W uses $N=2400000$ and $n=12289$, where the initial value of $h$ is set by the
relation $h\approx 7.23n^{-0.33}$ (see Fig. 5). Except for small residual differences
in the evolution of the maximum density at earlier collapse times, models U2W and
U4W shows essentially the same morphology. The same is true for models U1W and U3W.
It is important to notice that increasing $n$ with resolution implies reducing the
particle discretization errors {\bf and improving the mass resolution as
$M_{\rm min}=nm\sim n^{-1}$}. In other words, this means that the particle
approximation approaches the kernel approximation. Since models U1C--U4C work with
smaller smoothing lengths due to their fixed, low value of $n$ compared to models
U1W--U4W for the same $N$, it is not possible to establish a direct quantitative
comparison between both sequences. Indeed this will require recalculating models
U1W--U4W with huge amounts of neighbors so that both sequences will have the same
value of $h$ but different $N$. However, working with finer values of
$h$, while losing complete consistency, does not imply higher accuracy and convergence.
If $C^{0}$ and $C^{1}$ consistencies are achieved, it follows that the discrete
expansions (16) and (17) tend to their continuous counterparts (13) and (14),
respectively, implying second-order accuracy for the particle approximation 
independently of the numerical value of $h$. This is the essence of particle consistency 
in SPH.

The early collapse is qualitatively similar to models U1C--U4C. Initially the cloud
flattens about the equatorial plane, producing an isothermal disk with strong shocks
on both sides of it. The azimuthal structure of the
disk consists of two overdense blobs as a result of the $m=2$ perturbation seed,
which then fall toward the cloud center to merge into a bar with maximum density at
its endpoints. Due to converging flow into the bar, it soon grows in mass and
undergoes a cylindrical collapse upon itself. The result of this process
is the formation of a binary connected by a considerably more massive and thicker bar 
compared to models U1C--U4C. The basic features of the formation of the binary plus
connecting bar are very similar in Figures 8 and 9 despite the difference in spatial
resolution. In these models the bar is centrally condensed, a feature which is not
clear from models U1C--U4C. The bar as a whole is never seen to contract into a
singular filament. The nascent binary cores are spinning about an axis of
symmetry passing through their points of maximum density. This causes the bar to fan
out close to the fragments and develop well-pronounced spiral arms. As the cores 
accrete low angular momentum from the connecting bar and the spiral arms, the
binary separation decreases and the bar eventually dissipates. Because of its higher
initial resolution, model U4W fragments into a wider binary ($t=1.2869t_{\rm ff}$)
compared to model U2W ($t=1.2973t_{\rm ff}$). As a result of the accretion process,
well-defined protostellar disks form around the cores. The size of these disks is of
order $\sim 50$ AU. We note that the outcome of model U2W is very similar to that
reported by \cite{Klein99} for the same initial conditions using their AMR finite-difference
method. The last snapshot of Figure 8 shows the binary at an orbital separation
of $\sim 88$ AU, when almost 10\% of the cloud mass is cointained by the fragments.
A similar binary system is produced by models U3W and U4W but with larger orbital separations
($\sim 146$ AU) compared to model U2W at approximately the same maximum density.
However, in model U4W the circumstellar disk of one of the binary cores is seen to
fragment into a secondary of mass $\sim 0.02M_{\odot}$, which then revolves around the
primary with mean orbital separations of $\sim $14--20 AU. The last snapshot of Figure
9 shows the final configuration for model U4W, where a new small fragment
($\sim 0.006M_{\odot}$) has emerged from the residual bar material, which moves 
toward the binary core on the right side and so it will probably merge. The calculation
was terminated at this time because of the increasingly small time steps at this stage
of the evolution. About 9\% of the total cloud mass is contained by the cores in
models U3W and U4W. The formation of an apparently stable triple system in the highest
resolved calculation shows that the standard isothermal test is a demanding one.

Fragment disruption is never seen to occur and very good convergence is achieved.
This is a big difference with models U1C--U4C, where the cores disrupted into smaller
fragments and the connecting bar experienced multiple fragmentation along its length
into similar small fragments. {\bf The use of a Wendland function with a large number
of neighbors provides sufficient sampling of the kernel volumes and reduces particle noise
compared to the case of models U1C--U4C. Therefore, fragmentation of the protostellar disk
leading to a close binary in model U4W is not the result of noise amplification but 
rather of the nonlinear growth of a gravitational instability as the mass resolution
is improved (see Section 5.4 below)}. The effects of increasing the number of neighbors 
from $n=30$ to 200, while keeping $N$ fixed were previously studied by \cite{Commercon08}
for initially uniform clouds starting with stronger thermal support ($\alpha =0.50$)
and lower rotation ($\beta =0.04$) than the models considered here. They found that
increasing the ratio $n/N$ speeds up fragmentation because increasing $n$ for a fixed
$N$ decreases the spatial resolution as $h$ necessarily increases. We note that this
strategy is different from the one implemented here where full SPH consistency
demands that $n/N\to 0$ and $h\to 0$ in the limit $N\to\infty$ and $n\to\infty$
\citep{Rasio00}. The impact of varying the initial temperature on the collapse of the
standard isothermal test case was recently studied by \cite{Riaz14} using their
GRADSPH code. In particular, when the temperature is set to $T=10$ K they obtain
a stable binary system in a similar way as shown in Figure 8 for model U2W. However,
their calculations differ from ours in the value of $\rho _{\rm crit}$. {\bf If
$\rho _{\rm crit}$ is two orders of magnitude higher, this surely lengthens the
isothermal phase of collapse and favors} the formation of a stable binary system
(see, for instance, \cite{Arreaga07} for similar calculations with the standard 
GADGET-2 code; their Figures 5 and 6). The effects of the magnetic field on a variant
of the standard isothermal test case have been investigated by \cite{Burzle11} using
the development version of GADGET-3 extended to include the magnetic field. Setting
$\rho _{\rm crit}=1.0\times 10^{-14}$ g cm$^{-3}$, they also obtained a stable binary
system as in \cite{Arreaga07} and \cite{Riaz14} for a purely hydrodynamical calculation
with no magnetic field.

Figures 10 and 11 show the time evolution of the estimates of the moments for models U2W and
U4W. Compared to Figure 7, the moments $M_{0}$ and $\langle {\bf M}_{1}^{\prime}\rangle$
are now closer to unity for most of the evolution, implying that approximate $C^{0}$
consistency is achieved in this set of calculations, except after $\approx 1.26t_{\rm ff}$
when the degree of $C^{0}$ consistency is temporally lost within the fragment regions (see Fig. 
10 for model U2W). This occurs precisely when $h$ changes rapidly to ensure sufficient
spatial resolution within the higher-density regions. After this adaptive process, i.e., when
the variations of $h$ slow down, $C^{0}$ consistency is rapidly restored (after about
$1.34t_{\rm ff}$ for model U2W and $1.31t_{\rm ff}$ for model U4W). This is not surprising
since it is well-known that adaptive SPH calculations severely affect the consistency of
the method \citep{Liu06}. However, this temporal loss of consistency can be cured by
increasing further both $n$ and $N$ such that the ratio $n/N\to 0$. This can be seen by
comparing Figures 10 and 11, where the interval of inconsistency is reduced and the
quality of the estimates improves for model U4W. If we take the temporal mean of the maximum
of the distributions of $M_{0}$ and $\langle {\bf M}_{1}^{\prime}\rangle$ over the full
interval, the result is very close to unity, implying that $C^{0}$ consistency is
maintained on average. This is not the case for models U1C--U4C. The maximum of the
distributions for the other moments in Figures 10 and 11 are seen to exhibit erratic
oscillations about a mean value close to zero. However, the amplitudes of the oscillations
are much smaller for model U4W than for model U2W, implying that approximate $C^{1}$
consistency is better achieved by the former model. Therefore, we may conclude that 
model U4W is actually closer to second-order accuracy {\bf and exhibits less noise than
its counterpart models at lower resolution}.

\subsection{Collapse of the Gaussian cloud}

Calculations of the protostellar collapse starting from centrally condensed, Gaussian
density variations are of greater interest to understand the process of binary
fragmentation. A sequence of models similar to G1C--G6C with increasing spatial
resolution and fixed $n (=64)$ was previously calculated by \cite{Arreaga07} using
the standard GADGET-2 code (their models G1B--G6B). As the resolution was progressively
increased from $N=0.6$ to 10 million particles, they obtained apparent convergence to a binary
system. In order to separate the effects of the artificial viscosity from those of improved
consistency on fragmentation, we have run this set of models using the standard
GADGET-2 code with $n$ fixed to 64 neighbors and our improved artificial viscosity method
for approximately the same range of resolutions explored by \cite{Arreaga07} (see Table 2).
In this case, the sequence of calculations G1C--G6C all produced triple systems,
consisting of a bound binary plus an ejected third fragment escaping to infinity for 
models G1C and G2C. In contrast, models G3C--G6C also produced a bound binary with
the third fragment now orbiting around the binary core at distances from $\sim 4$ to 5 times
larger than the binary separation. Although the outcome is the same for all models, the
details of the final patterns and properties of the fragments are not the same
implying a lack of convergence which can be associated with a loss of $C^{0}$
consistency as revealed by the time evolution of the distributions of the estimates of 
$M_{0}$ and $\langle {\bf M}_{1}^{\prime}\rangle$. As was outlined before, the form
of the artificial viscosity affects the outcome of the simulations. Unphysical 
dissipation of local velocity gradients away from shocks in the calculations of
\cite{Arreaga07} is likely to be the cause of the differences with the outcomes of sequence 
G1C--G6C.

We now describe the results of models G1W--G6W, which were run using the modified
GADGET-2 code with increasing number of neighbors. The initial phase of collapse for
these models is qualitatively similar to that observed for models G1C--G6C. That is,
up to the point where $\rho _{\rm max}=\rho _{\rm crit}$, the cloud evolves to a
centrally condensed, flat disk. When the disk becomes adiabatic, it expands due to
increasing pressure forces and deforms by rotational effects into a central bar.
Because of further rotation, the bar {\bf wraps} up and becomes ${\rm S}$-shaped. After
about a rotation period of the central bar, the ${\rm S}$-shaped structure grows in
size and develops long arms. Meantime, the bar continues rotating and collapses into
a central blob. By this time, the ${\rm S}$-shaped structure has already deformed and
two satellite fragments form at the end parts of the winding arms and at the same
distance from the central core, giving rise to a
ternary core. From top to down, Figure 12 shows column density images of the evolution
of models G3W--G6W at comparable maximum densities and same times. It is evident from the
first and second column of images that fragmentation is anticipated when the resolution
is increased. However, from the last column we may see that reasonably good convergence
is achieved by the highly resolved calculations at comparable maximum
densities and times.  By $2.0192t_{\rm ff}$, the fragments are
well-defined and evolving as separate entities from the parent cloud. They
contain about 13.2\% (G3W), 8.4\% (G4W), 8.5\% (G5W), and 8.5\% (G6W) of the total cloud
mass, while the separations of the two satellite fragments from the central core are
($\sim 298$ AU) for G3W, ($\sim 279$ AU) for G4W, ($\sim 289$ AU) for G5W, and
(288 AU) for G6W.

Figures 13 and 14 depict the time evolution of the distribution of the estimates of the
moments for models G3W and G6W. As for the standard isothermal case, only the late evolution
is represented in both figures and the distributions are constructed by considering only
particles with densities $>\rho _{\rm crit}$. Approximate $C^{0}$ and $C^{1}$ consistencies
are achieved in both cases. However, by comparing these two figures we may see that the
quality of the simulation improves for model G6W working with higher values of $n$ and $N$.
Therefore, as the values of $n$ and $N$ are increased, the particle discretization errors
decay asymptotically and the calculations become closer
to second-order accuracy. As was stated by \cite{Commercon08}, in studies of protostellar
collapse and fragmentation it is more difficult to attain convergence for low than for
high thermal support. The point is that in the case of low thermal support the dynamics
of the flow is likely to become highly nonlinear faster than in clouds with high
thermal support. The same is also true for models where the isothermal phase of collapse is 
prolonged by choosing high values of the critical density, as is indeed the case of
models U1W--U4W, where $\rho _{\rm crit}=5.0\times 10^{-12}$ g cm$^{-3}$. According to
expansions (16) and (17), if $C^{1}$ particle consistency is restored, the errors carried
by the estimates of a function and its gradient match those for the kernel approximation
($\sim h^{2}$). However, as $n$ and $N$ are increased, $h$ decreases (see Figure 5). 
Thus, decreasing the size of the kernel not only improves the resolution but also
favors the growth of nonlinearity at smaller scales. If thermal support is retarded, then
nonlinear behavior may amplify and lead to further fragmentation. This is precisely the
difference between the outcomes of models U1W--U3W (Figure 8) and model U4W
(Figure 9), where further fragmentation is observed. This is not the case in Figure 12,
where the transition from isothermal to adiabatic collapse is anticipated, and the
higher resolution calculations provide almost the same fragmentation time and pattern.

\subsection{Protostellar disk fragmentation}

{\bf The mass contained within the kernel volume scales with $h$ as $\sim h^{3}$.
Therefore, if $h\sim n^{-1/3}$ then the minimum mass varies with $n$ as 
$M_{\rm min}\sim n^{-1}$, implying that large numbers of neighbors leads to improved  
mass resolution. This aspect makes a big difference with models U1C-U4C and G1C-G6C,
which employ a fixed value of $n (=64)$ regardless of the total number of particles.
Improving the mass resolution will certainly allow to better resolve small-scale 
features in the flow during the collapse and fragmentation of protostellar cloud cores. 
This is the case of the highly resolved models U4W and G4W-G6W, where after large-scale
fragmentation, which was seeded here by a background $m=2$ density variation,
well-defined rotating disks were clearly seen to form around the growing fragments
as a result of infalling material from the cloud envelope. In particular, Fig. 15 shows
enlarged density maps for the evolution of one of the former binary fragments formed in
model U4W (see Fig. 9, leftmost fragment at $t=1.3252t_{\rm ff}$). As the fragment grows
in density, a circumstellar disk forms which then becomes sufficiently massive to develop
a two-armed spiral structure associated with the linear growth stage of a gravitational 
instability. By this time ($1.3042t_{\rm ff}$), the mass of the disk is 
$\approx 0.011$ $M_{\odot}$ compared to $\approx 0.032$ $M_{\odot}$ of the central protostar.
The radius of the disk is $R_{\rm disk}\approx 25$ AU and grows to $\approx 36$ AU by
$t=1.3109t_{\rm ff}$ just before fragmentation. According to the Toomre stability
criterion, the disk is unstable to axisymmetric perturbations if
\begin{equation}
Q\approx \frac{2M_{\star}}{M_{\rm disk}}\frac{H}{r}<1,
\end{equation}
where $M_{\star}$ is the mass of the central protostar, $H$ is the disk scale height,
and $r$ denotes radial distance from the central protostar. In the above definition 
we have assumed that the disk is Keplerian and define $H=c/\omega$, where $\omega$ is the
Keplerian angular velocity at radius $r$. Taking $r=20$ AU, which is the approximate 
radius where fragmentation of the disk occurs, we find that $H/r\approx 0.17$ and the 
Toomre parameter $Q\approx 0.97$. Soon thereafter, the gravitational
instability enters a nonlinear growth phase and the outermost part of one of the arms
condenses into a secondary at a distance of $\approx 18$ AU from the primary, leading 
to fragmentation of the disk into a tight binary. The newly formed fragment takes its orbital
angular momentum from the rotation of the disk and revolves around the primary in an
approximate circular orbit. By $1.327t_{\rm ff}$, when the calculation is terminated,
the primary has a mass of $\approx 0.044$ $M_{\odot}$, while the mass of the secondary is
$\approx 0.02$ $M_{\odot}$.

The calculation of model U4W shows that working with 12289 neighbors is enough to
resolve small-scale fragmentation due to the gravitational instability of a massive
protostellar disk. While this can be seen as a possible mechanism for the formation
of binary/multiple stellar systems separated by a few AU, recent observations of
the L1448 IRS3B triple system, consisting of two protostars at the center and a third
one distant from them, are providing strong support to this conclusion \citep{Tobin16}.
The observations, which were conducted with the Atacama Large Millimeter/submillimeter
Array (ALMA), show that the spiral structure in the dusty disk surrounding the young stars
is indicative of their having been formed by fragmentation of the disk via a gravitational
instability. Hence, differences in distance may be the result of different formation
mechanisms. For instance, systems separated by hundreds to thousands AU are likely to
be the result of fragmentation of the larger cloud during its early gravitational
collapse, while tighter systems with separations of tens of AU may be hierarchical
systems formed from disk fragmentation.

Similarly, as shown in Fig. 12, models G1W-G6W collapsed to form a central protostar
surrounded by a circumstellar disk, which then experienced fragmentation into two
secondaries, forming a tertiary protostellar system. This time the circumstellar disk
appears to be larger ($\gtrsim 600$ AU) at the time of fragmentation compared to
model U4W because of thermal retardation due to the assumption of a lower value of
the critical density ($=5.0\times 10^{-14}$ g cm$^{-3}$) for the Gaussian models. As
the ternary fragments grow in density, each of them develop well-pronounced 
circumstellar disks as shown by the last column of density maps in Fig. 12 for
models G4W-G6W}.

\section{Conclusions} \label{sec: conclusions}

We have investigated the consistency of smoothed particle hydrodynamics (SPH) in
numerical simulation tests of the collapse and fragmentation of rotating molecular cloud
cores. A modified version of the simulation code GADGET-2 \citep{Springel05} was used for
the calculations, where the interpolation kernel was replaced by a Wendland C$^{4}$ function 
to allow support of large numbers of neighbors and an advanced scheme for the artificial
viscosity was implemented based on the method proposed by \cite{Hu14}. Approximations
to the power-law relations provided by \cite{Zhu15} were used to set the kernel 
interpolation parameters, namely the total number of particles $N$, the smoothing length
$h$, and the number of neighbors $n$, where $h$ is allowed to vary with $N$ as 
$h\sim N^{-1/6}$. With this choice, the scalings $h\approx 7.23n^{-0.33}$ and
$n\approx 7.61N^{0.503}$ were used to set the initial values of $h$ and $n$ for fixed $N$.
As the domain resolution is increased, these scalings comply with the combined limit
$N\to\infty$, $h\to 0$, and $n\to\infty$ with $n/N\to 0$ for full SPH consistency
\citep{Rasio00}.
 
The initial conditions for the protostellar collapse models were chosen to be the
``standard isothermal test case'' in the variant calculated by \cite{Burkert93} and
the centrally condensed, Gaussian cloud advanced by \cite{Boss91}, coupled to a barotropic
pressure-density relation to simulate the transition from the isothermal to the nonisothermal 
collapse. The critical density, separating both regimes, was set to
$\rho _{\rm crit}=5.0\times 10^{-12}$ g cm$^{-3}$ for the standard isothermal test
to provide insight into the role played by $n$ for a case where convergence is
more demanding at late stages of the evolution. In contrast, for the
Gaussian cloud model $\rho _{\rm crit}=5.0\times 10^{-14}$ g cm$^{-3}$, which is more
representative of the near-isothermal phase \citep{Boss00}. Since convergence is easier
when shortening the isothermal phase of collapse due to thermal retardation, this model
has been used to discern the effects of the artificial viscosity on the final outcome
by comparing with previous calculations by \cite{Arreaga07}.

Two separate sequences of calculations with increasing $N$ were run for both models. One
sequence used the standard version of GADGET-2 with fixed $n(=64)$, while the other
sequence was calculated using the modified version of the code with varied $n$.
Over $\sim 9$ orders of magnitude
increase in density, the standard isothermal models with fixed $n$ produced a binary
connected by a singular bar in much the same way as reported in previous SPH calculations.
However, as the evolution was continued farther in time the binary cores and the bar
were seen to fragment into smaller condensations, with the dynamics becoming highly nonlinear 
and chaotic due to numerical noise amplification. At these stages, even the highly
resolved models showed no sign of convergence. In contrast, the models with varied $n$
experienced a similar initial collapse, producing stable binary systems for moderate
resolutions ($N\leq 1200000$) and a triple system for $N=2400000$ due to fragmentation
of the disk around one of the binary components into a secondary.
{\bf Owing to the higher number of neighbors
and hence improved mass resolution for this model, it was possible to resolve the
small-scale structure and fragmentation of the disk into a close binary. This 
mechanism has recently received convincing observational evidence for explaining the
formation of close binary/multiple protostellar systems \citep{Tobin16}}.  
On the other hand, the Gaussian
clouds using the standard GADGET-2 code with fixed $n$ but with the new scheme of the
artificial viscosity produced different outcomes at all resolutions compared to those
previously reported by \cite{Arreaga07}. In all cases, only qualitative convergence into
a triple system was achieved, consisting of a bound binary plus a third core at much
higher orbital distances. Evidently, the reduced dissipation and better treatment of
shocks implied by the new artificial viscosity had an important impact on the final
outcome. With the modified code, all runs also produced a final triple system but with
a quite different pattern. In this case, the highly resolved runs were seen to converge
at comparable maximum densities and times.

The degree of consistency of the calculations was measured by tracking how well the
kernel consistency relations were reproduced by the particle
approximation. From the time evolution of the estimates of the moments of the kernel
it was clear that all calculations with fixed $n(=64)$ were inconsistent. The normalization
condition of the kernel and the first moment of the gradient always diverged from
unity, with the maximum deviations occurring at the late stages of the evolution just
after the fragmentation period when the fragments were growing in density and $h$ was
varying rapidly to ensure adequate spatial resolution, meaning that $C^{0}$ consistency
was not achieved by these models. Thus, violation of the normalization condition by the
particle approximation implies that these calculations are even {\bf worse} than first-order
accurate due to persisting zeroth-order discretization errors \citep{Read10}. In
contrast, approximate $C^{0}$ and $C^{1}$ consistencies were achieved by all models
when $n$ was allowed to vary with $N$. However, loss of $C^{0}$ consistency was
temporally observed within the fragment regions due to rapid variations of $h$ there,
confirming previous expectations that adaptive kernels
severely affect the consistency of SPH \citep{Liu06}. After this adaptive process, the
variations of $h$ in the high-density regions slowed down and $C^{0}$ consistency was
rapidly restored by the models. In both sequences of calculations, as $n$ and $N$ are
increased the interval and degree of inconsistency are progressively reduced and the
quality of the calculations is improved. On the other hand, the temporal means of the
estimates of the normalization condition of the kernel and the first moment of the
gradient over the full interval, where consistency is lost and then restored, are seen
to peak very close to unity, implying that $C^{0}$ consistency is achieved on average
within the fragment regions. Since the estimates of the second moments are always close
to zero, approximate $C^{1}$ consistency is also achieved. We may therefore conclude that
the simulations presented here are actually close to second-order accuracy.

As a final remark, it has been demonstrated that $C^{0}$ particle consistency for both
the estimates of a function and its gradient implies preservation of the homogeneity
and isotropy of the discrete space, which have as consequences conservation of the
linear and angular momentum, respectively \citep{Vignjevic09,Sigalotti16}. Therefore,
we may expect that local linear and angular momentum are well conserved in our consistent
collapse calculations. However, it would be interesting to quantify numerically the
degree of angular momentum conservation when $C^{0}$ consistency is achieved in the
limit $n/N\to 0$. Due to its Lagrangian character, SPH provides direct access to the
initial angular momentum of particles so that any loss can be easily quantified following
a similar analysis to that developed by \cite{Commercon08}. Future studies in this
line will deal with the impact of consistency on angular momentum conservation and how
to address the Jeans-resolution requirement under large numbers of neighbors. 

\acknowledgments

{\bf We thank the anonymous referee for raising a number of comments and suggestions that
have improved the style and content of the manuscript. In particular, his/her
comment on the relation between consistency and mass resolution is much acknowledged}.
The calculations of this paper were performed using the computing facilities of
ABACUS-Centro de Matem\'aticas Aplicadas y C\'omputo de Alto Rendimiento of Cinvestav-IPN.
This work was partially supported by ABACUS through the CONACyT grant EDOMEX-2011-C01-165873
and by the Departamento de Ciencias B\'asicas e Ingenier\'{\i}a (CBI) of the Universidad
Aut\'onoma Metropolitana--Azcapotzalco (UAM-A), the Instituto de Ciencias B\'asicas e
Ingenier\'{\i}as of the Universidad Aut\'onoma del Estado de Hidalgo (UAEH), and the
Instituto Venezolano de Investigaciones Cient\'{\i}ficas (IVIC) through internal funds.

\clearpage

\begin{deluxetable}{lcccccccc}
\tablenum{1}
\tablecaption{Standard deviation $\sigma (\cdot)$ and expectation value $\langle\cdot\rangle _{e}$
of the density and moments estimates as a function of $n$.}
\tablewidth{0pt}
\tablehead{}
\decimalcolnumbers
\startdata
Glass & & & & & & & & \\
$n$ & 48 & 64 & 120 & 240 & 480 & 800 & 1600 & 3200 \\
$\sigma (\rho)$ & 3.44(-2) & 1.67(-2) & 3.31(-3) & 1.31(-3) & 8.85(-4) & 7.03(-4) & 4.85(-4) & 3.12(-4) \\
$\langle\rho\rangle _{e}$ & 1.05479 & 1.02380 & 1.00421 & 1.00064 & 1.00003 & 0.99994 & 0.99990 & 0.99989 \\
$\sigma (M_{0})$ & 1.93(-2) & 1.04(-2) & 2.46(-3)  & 7.36(-4) & 3.89(-4) & 3.51(-4) & 2.72(-4) & 1.87(-4) \\
$\langle M_{0}\rangle _{e}$ & 0.570 & 0.678 & 0.828 & 0.914 & 0.957 & 0.974 & 0.987 & 0.993 \\
$\sigma ({\bf M}_{1})$ & 2.15(-4) & 1.19(-4) & 1.75(-5) & 4.46 (-6) & 3.21(-6) & 3.58(-6) & 3.96(-6) & 3.76(-6) \\
$\langle {\bf M}_{1}\rangle _{e}$ & 0.0 & 0.0 & 0.0 & 0.0 & 0.0 & 0.0 & 0.0 & 0.0 \\
$\sigma ({\bf M}_{0}^{\prime})$ & 2.63 & 1.21 & 1.76(-1) & 3.55(-2) & 1.08(-2) & 7.97(-3) & 5.45(-3) & 3.18(-3) \\
$\langle {\bf M}_{0}^{\prime}\rangle _{e}$ & 0.0 & 0.0 & 0.0 & 0.0 & 0.0 & 0.0 & 0.0 & 0.0 \\
$\sigma ({\bf M}_{1}^{\prime})$ & 3.47(-2) & 2.31(-2) & 4.73(-3) & 9.18(-4) & 2.75(-4) & 1.75(-4) & 1.17(-4) & 8.71(-5) \\
$\langle {\bf M}_{1}^{\prime}\rangle _{e}$ & 0.86317 & 0.93636 & 0.98910 & 0.99792 & 0.99967 & 0.99991 & 0.99997 & 0.99999 \\
$\sigma ({\bf M}_{2}^{\prime})$ & 2.14(-4) & 1.56(-4) & 3.93(-5) & 6.63(-6) & 2.65(-6) & 2.38(-6) & 2.31(-6) & 2.05(-6) \\
$\langle {\bf M}_{2}^{\prime}\rangle _{e}$ & 0.0 & 0.0 & 0.0 & 0.0 & 0.0 & 0.0 & 0.0 & 0.0 \\
Random & & & & & & & & \\
$\sigma (\rho)$ & 2.97(-1) & 2.17(-1) & 1.20(-1) & 6.66(-2) & 3.75(-2) & 2.45(-2) & 1.37(-2) & 7.67(-3) \\
$\langle\rho\rangle _{e}$ & 1.22783 & 1.14226 & 1.05358 & 1.01846 & 1.00626 & 1.0027 & 1.00077 & 1.00019 \\
$\sigma (M_{0})$ & 1.03(-1) & 9.54(-2) & 6.73(-2) & 4.10(-2) & 2.39(-2) & 1.59(-2) & 9.04(-3) & 5.06(-3) \\
$\langle M_{0}\rangle _{e}$ & 0.541 & 0.659 & 0.821 & 0.912 & 0.956 & 0.974 & 0.987 & 0.993 \\
$\sigma ({\bf M}_{1})$ & 6.55(-4) & 6.23(-4) & 5.04(-4) & 3.74(-4) & 2.71(-4) & 2.13(-4) & 1.53(-4) & 1.09(-4) \\
$\langle {\bf M}_{1}\rangle _{e}$ & 0.0 & 0.0 & 0.0 & 0.0 & 0.0 & 0.0 & 0.0 & 0.0 \\
$\sigma ({\bf M}_{0}^{\prime})$ & 10.30 & 7.73 & 3.90 & 1.78 & 8.06(-1) & 4.49(-1) & 2.03(-1) & 9.13(-2) \\
$\langle {\bf M}_{0}^{\prime}\rangle _{e}$ & 0.0 & 0.0 & 0.0 & 0.0 & 0.0 & 0.0 & 0.0 & 0.0 \\
$\sigma ({\bf M}_{1}^{\prime})$ & 8.47(-2) & 6.82(-2) & 4.11(-2) & 2.43(-2) & 1.42(-2) & 9.49(-3) & 5.39(-3) & 3.05(-3) \\
$\langle {\bf M}_{1}^{\prime}\rangle _{e}$ & 0.695 & 0.794 & 0.915 & 0.969 & 0.989 & 0.995 & 0.998 & 0.999 \\
$\sigma ({\bf M}_{2}^{\prime})$ & 4.32(-4) & 3.89(-4) & 2.99(-4) & 2.24(-4) & 1.64(-4) & 1.29(-4) & 9.21(-5) & 6.61(-5) \\
$\langle {\bf M}_{2}^{\prime}\rangle _{e}$ & 0.0 & 0.0 & 0.0 & 0.0 & 0.0 & 0.0 & 0.0 & 0.0 
\enddata
\end{deluxetable}

\clearpage

\begin{deluxetable}{lclc}
\tablewidth{0pc}
\tablenum{2}
\tablecolumns{4}
\tablecaption{Collapse models\label{tab2}}
\tablehead{
\colhead{Model} & \colhead{Total number of} & \colhead{Number of} &
\colhead{Final} \\
\colhead{} & \colhead{particles ($N$)} & \colhead{neighbors ($n$)} &
\colhead{outcome} }
\startdata
  Uniform clouds & \\
\hline
  U1C............. & ~~~300,000 & ~~~64 & Binary? \\
  U2C............. & ~~~600,000 & ~~~64 & Binary? \\
  U3C............. & ~1,200,000 & ~~~64 & Binary? \\
  U4C............. & ~2,400,000 & ~~~64 & Binary? \\
\hline
  U1W............. & ~~~300,000 & ~4321 & Binary~ \\
  U2W............. & ~~~600,000 & ~6121 & Binary~ \\
  U3W............. & ~1,200,000 & ~8673 & Binary~ \\
  U4W............. & ~2,400,000 & 12289 & Triple~ \\ 
\hline 
  Gaussian clouds & \\
\hline
  G1C............. & ~~~300,000 & ~~~64 & Triple \\
  G2C............. & ~~~600,000 & ~~~64 & Triple \\
  G3C............. & ~1,200,000 & ~~~64 & Triple \\
  G4C............. & ~2,400,000 & ~~~64 & Triple \\
  G5C............. & ~4,800,000 & ~~~64 & Triple \\
  G6C............. & ~9,600,000 & ~~~64 & Triple \\
\hline
  G1W............. & ~~~300,000 & ~4321 & Triple \\
  G2W............. & ~~~600,000 & ~6121 & Triple \\
  G3W............. & ~1,200,000 & ~8673 & Triple \\
  G4W............. & ~2,400,000 & 12289 & Triple \\
  G5W............. & ~4,800,000 & 17412 & Triple \\
  G6W............. & ~9,600,000 & 24673 & Triple \\
\enddata
\end{deluxetable}

\clearpage

\begin{figure}
\figurenum{1}
\epsscale{1.6}
\plottwo{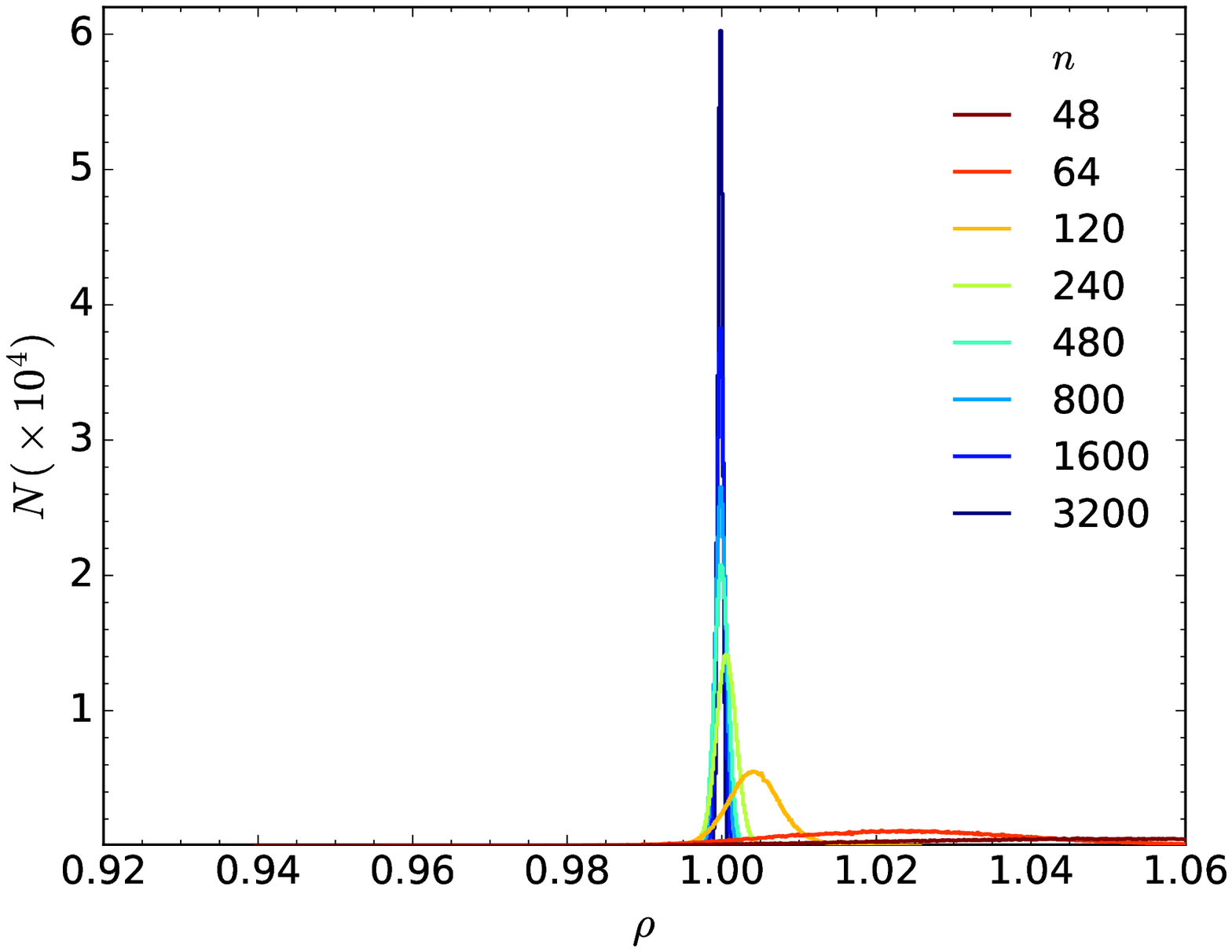}{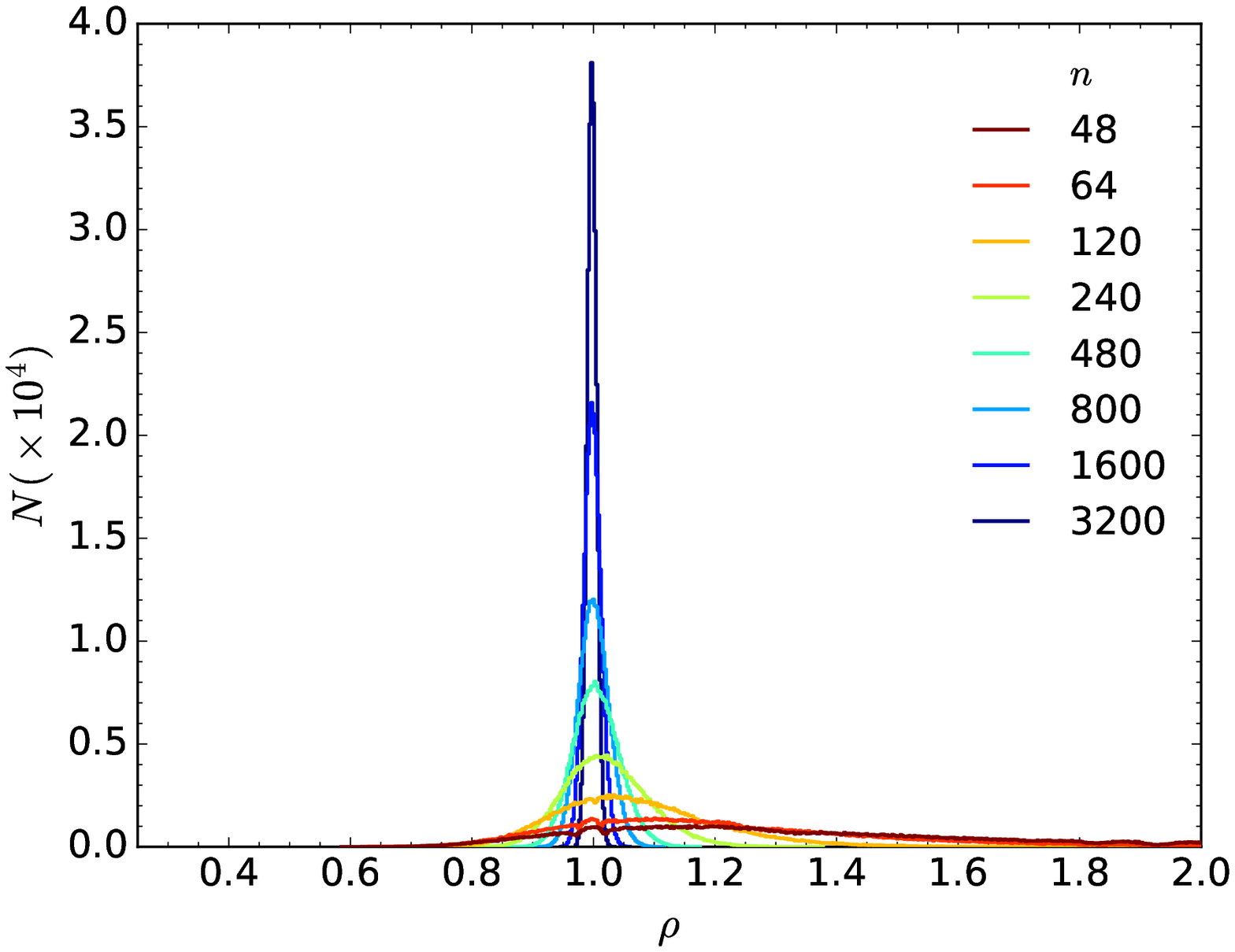}
\caption{Distribution of the particle density estimates as obtained using the
Wendland C$^{4}$ function with increasing number of neighbors from $n=48$ (red curve) to
3200 (dark-blue curve) for a glass ({\it top}) and a random ({\it bottom}) particle
distribution. In the former case, as $n$ is increased the density distribution approaches
a Dirac-$\delta$ distribution, while in the latter case, the density distribution looks
much broader and approaches a Gaussian-like distribution with a peak at $\rho =1$ for
$n\geq 800$. \label{fig:f1}}
\end{figure}

\clearpage

\begin{figure}
\epsscale{1.0}
\plottwo{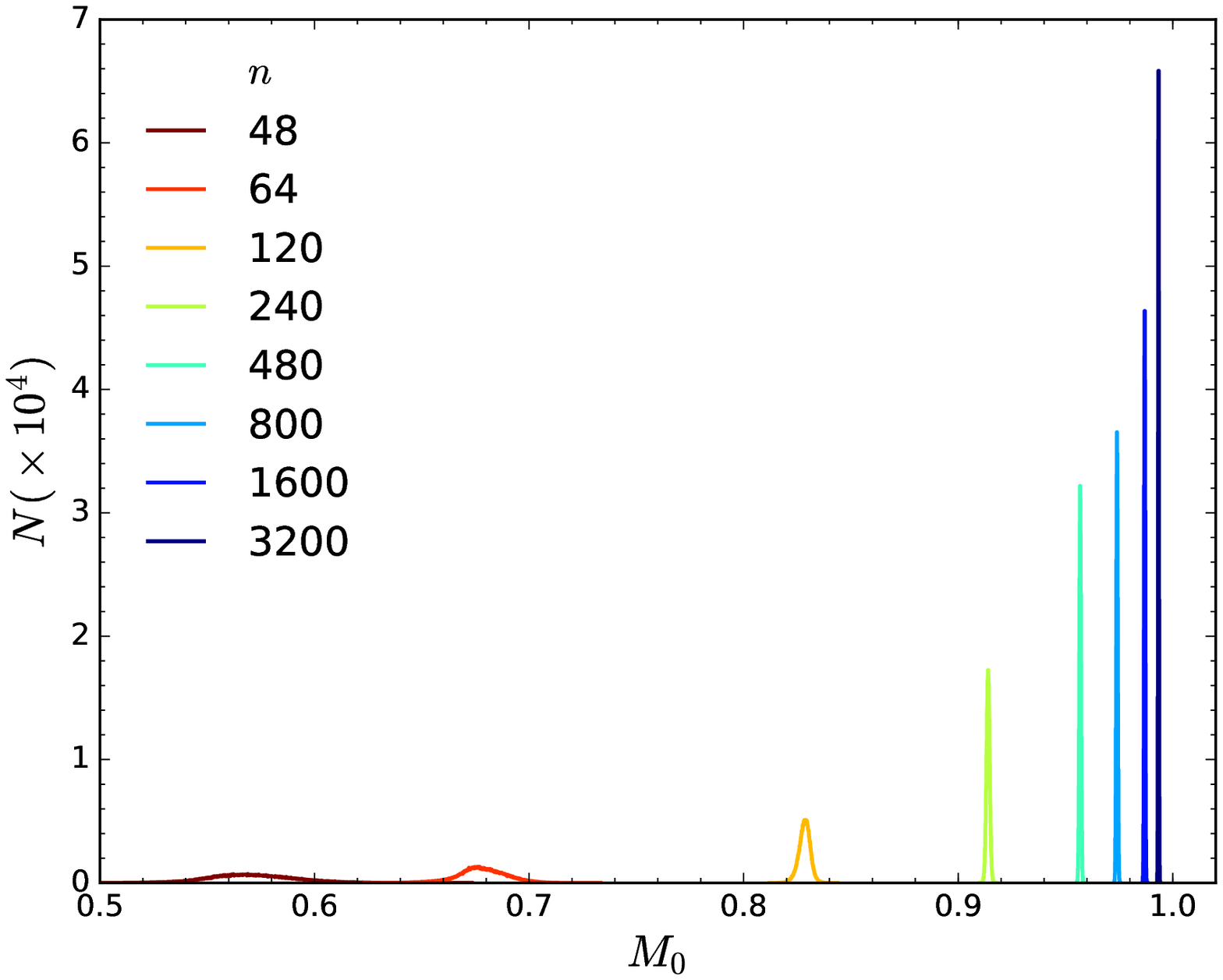}{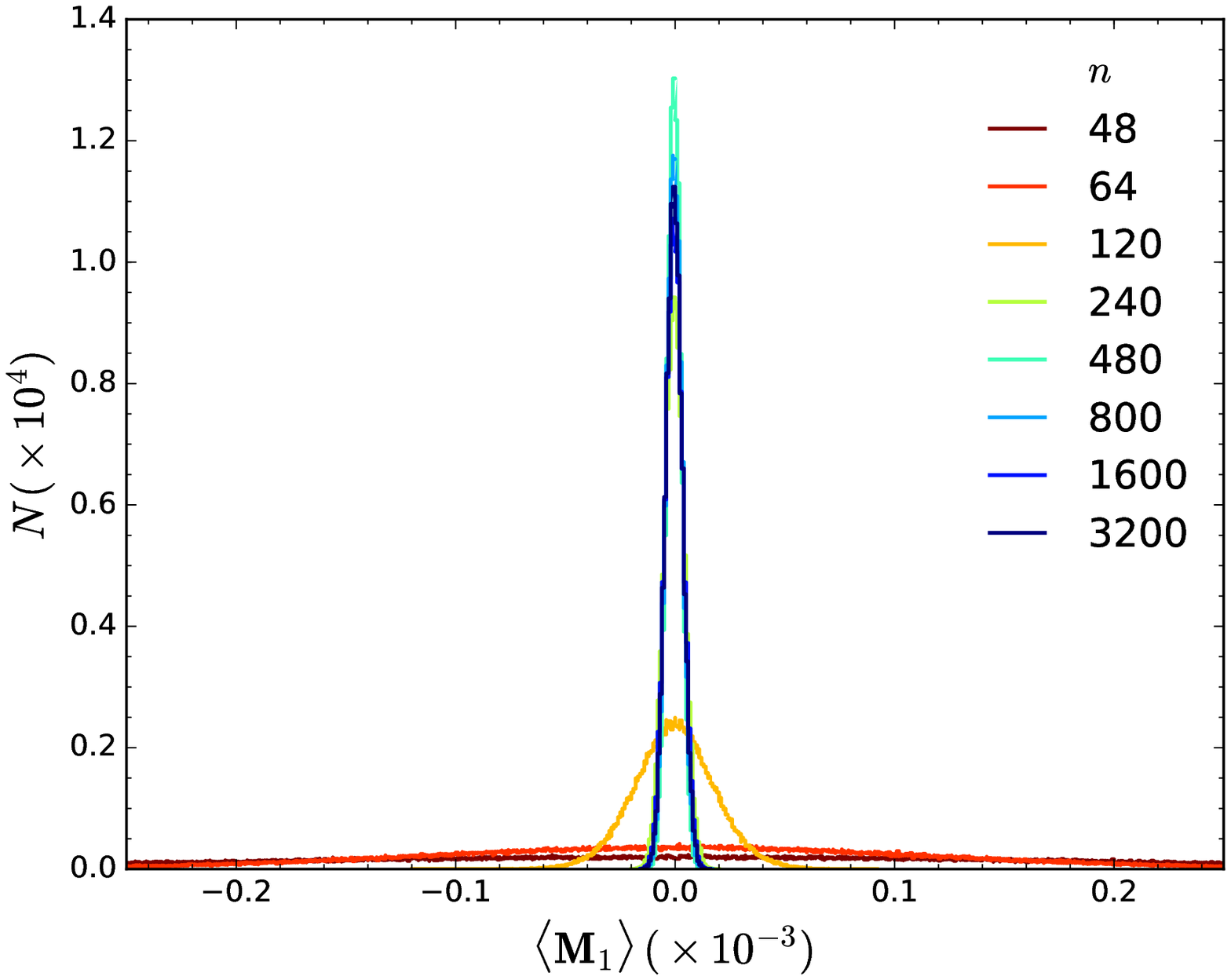}
\end{figure}
\begin{figure}
\epsscale{1.0}
\plottwo{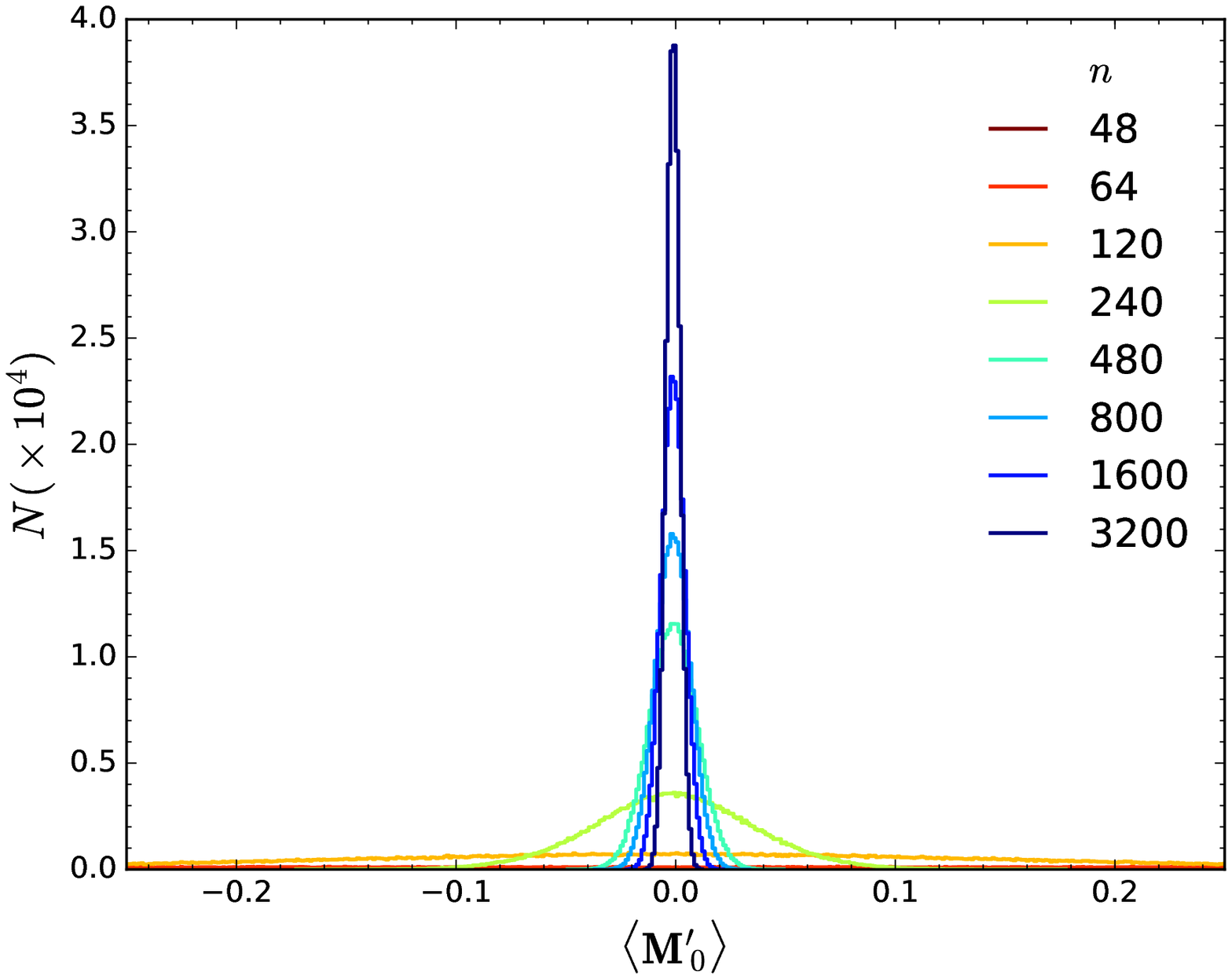}{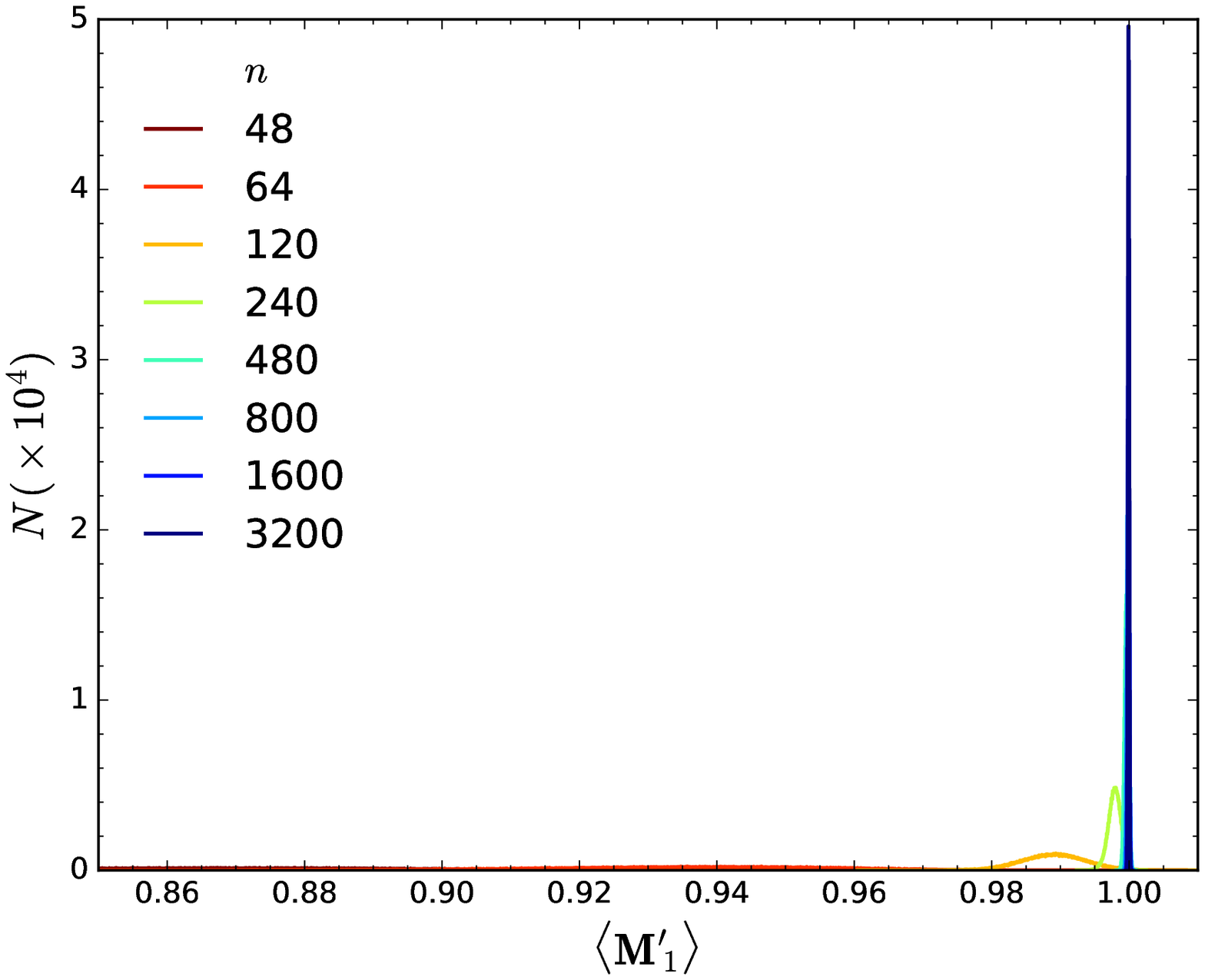}
\end{figure}
\begin{figure}
\figurenum{2}
\epsscale{0.5}
\plotone{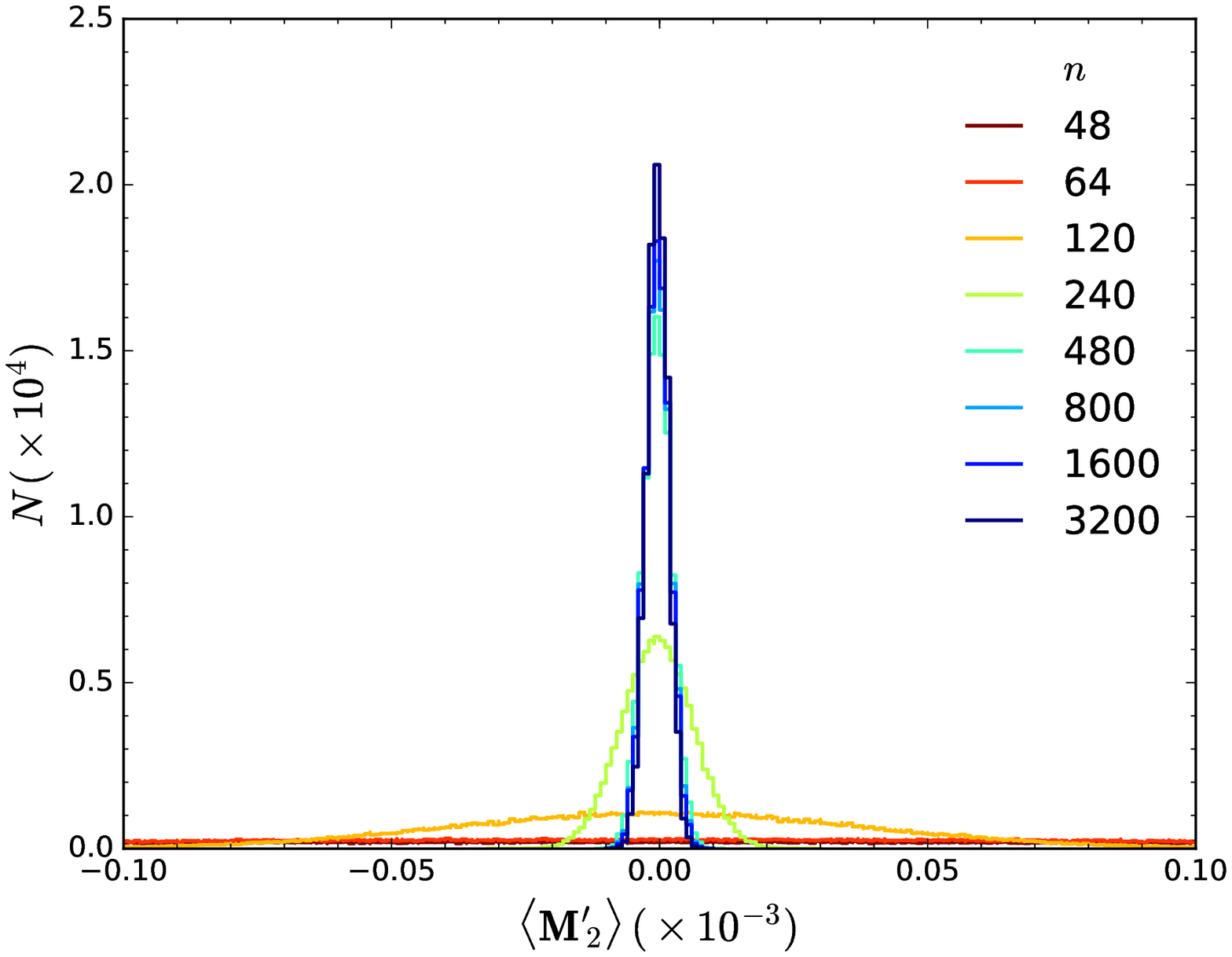}
\caption{Distributions of the moments of the kernel $M_{0}$ and ${\bf M}_{1}$ as given by
Eqs. (18) and (19) ({\it upper plots}) and the moments of the kernel gradient ${\bf M}_{0}^{\prime}$,
${\bf M}_{1}^{\prime}$, and ${\bf M}_{2}^{\prime}$ as given by Eqs. (20) and (21)
({\it middle and bottom plots}) for the glass particle distribution, using the Wendland C$^{4}$
function with increasing number of neighbors from $n=48$ (red curves) to 3200 (dark-blue curves).
$\langle {\bf M}_{1}\rangle$ and $\langle {\bf M}_{0}^{\prime}\rangle$ are the means of the
components of ${\bf M}_{1}$ and ${\bf M}_{0}^{\prime}$, while $\langle {\bf M}_{1}^{\prime}\rangle$ 
and $\langle {\bf M}_{2}^{\prime}\rangle$ correspond to the mean of the elements of matrices
${\bf M}_{1}^{\prime}$ and ${\bf M}_{2}^{\prime}$, respectively.
\label{fig:f2}}
\end{figure}

\clearpage

\begin{figure}
\epsscale{1.0}
\plottwo{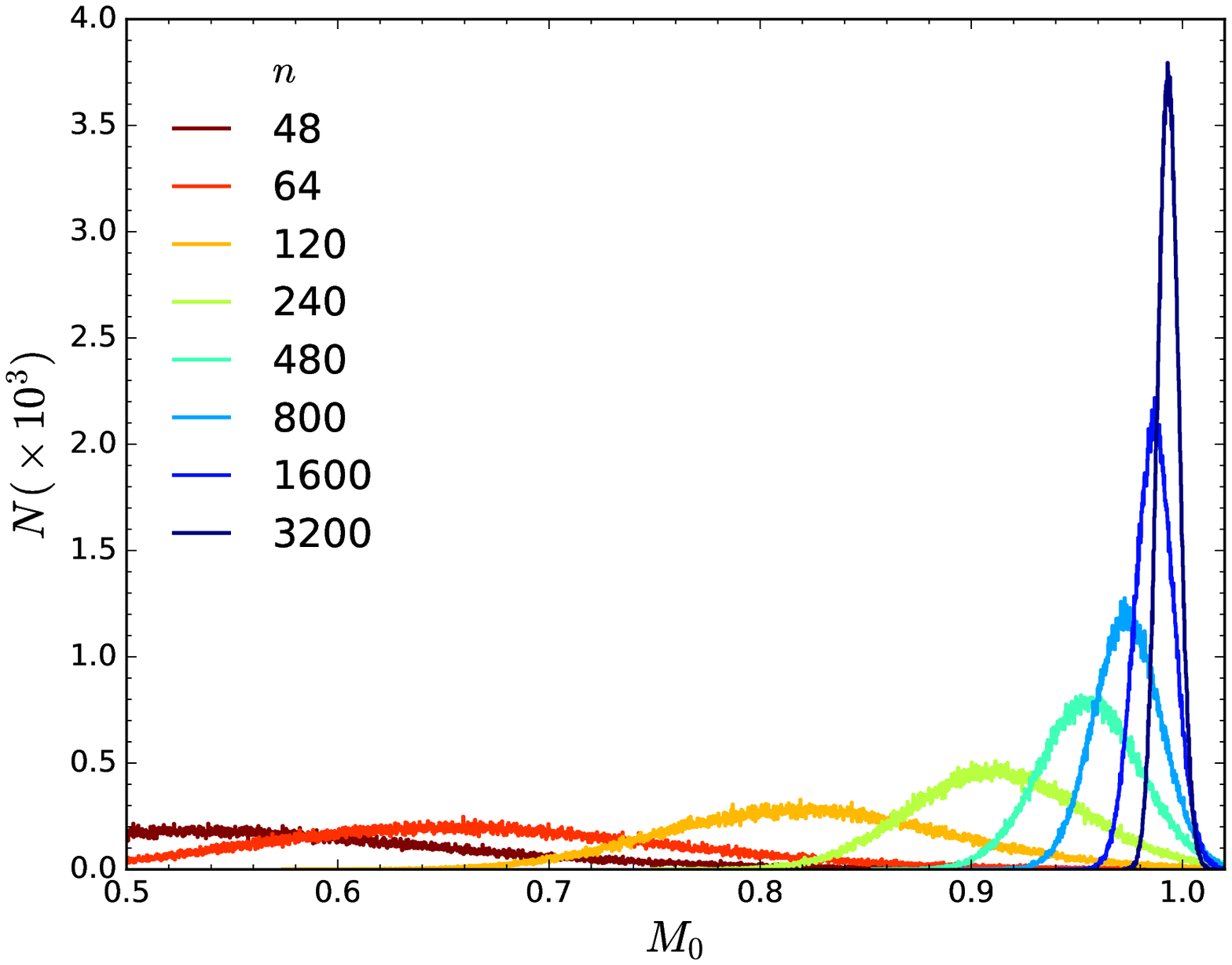}{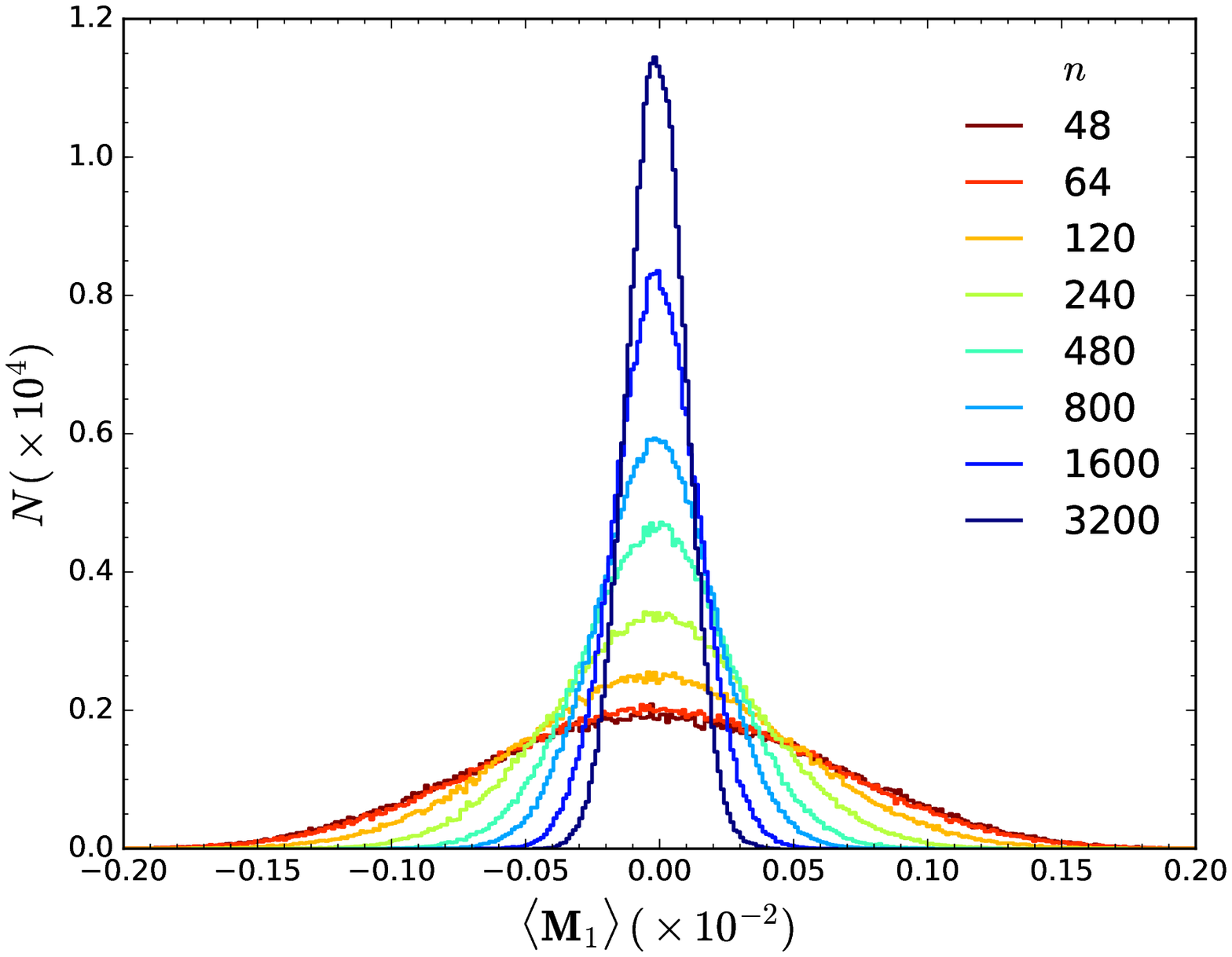}
\end{figure}
\begin{figure}
\epsscale{1.0}
\plottwo{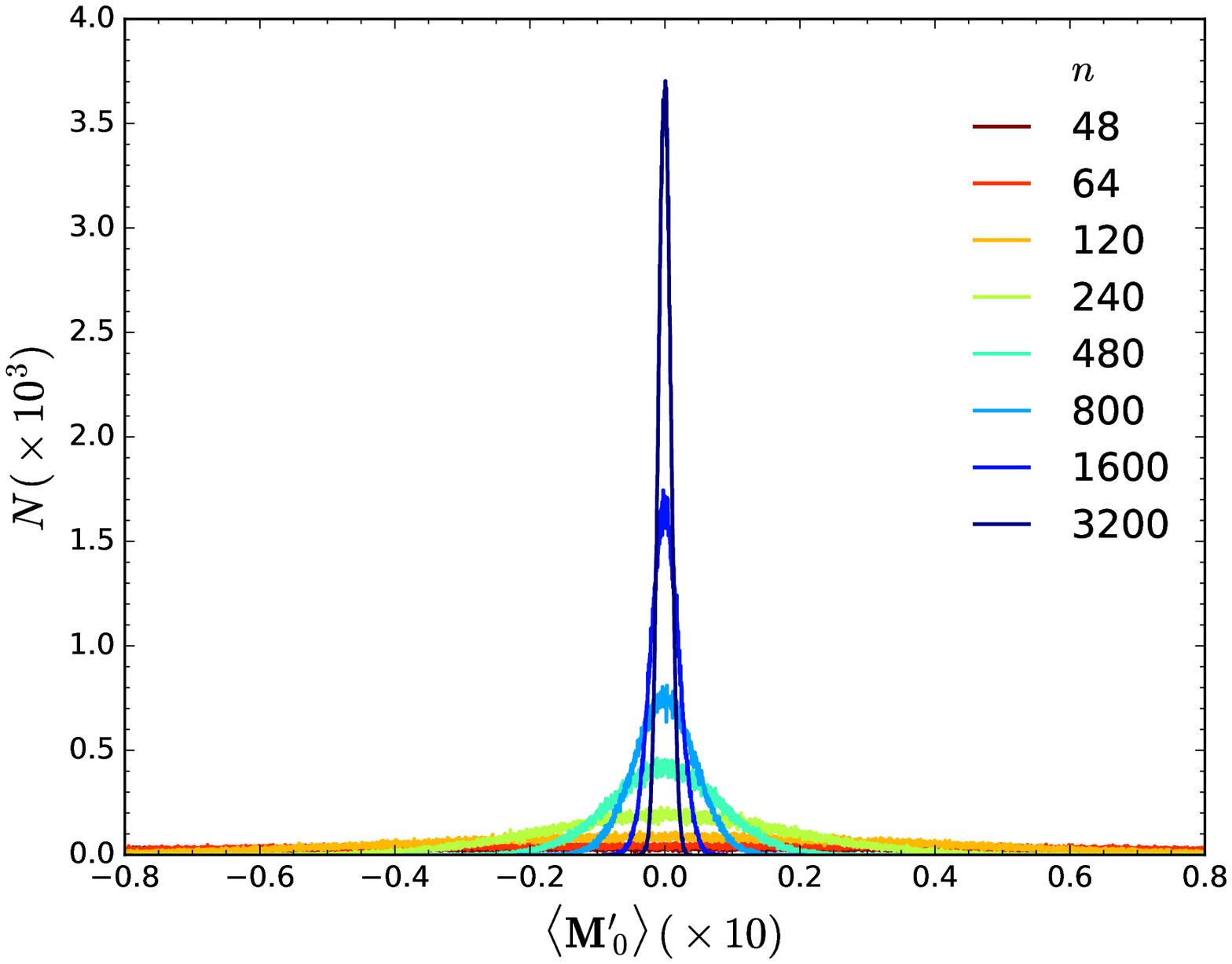}{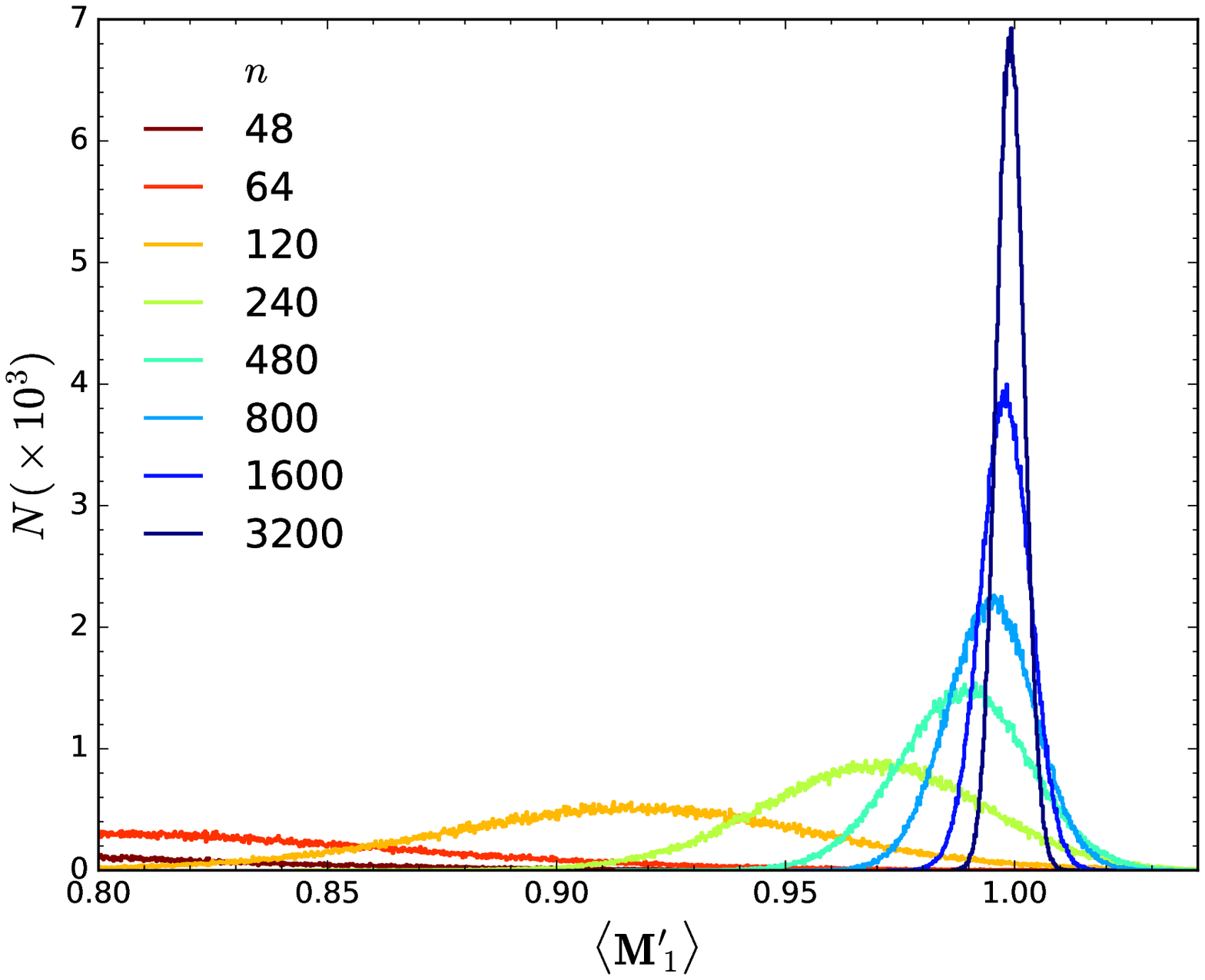}
\end{figure}
\begin{figure}
\figurenum{3}
\epsscale{0.5}
\plotone{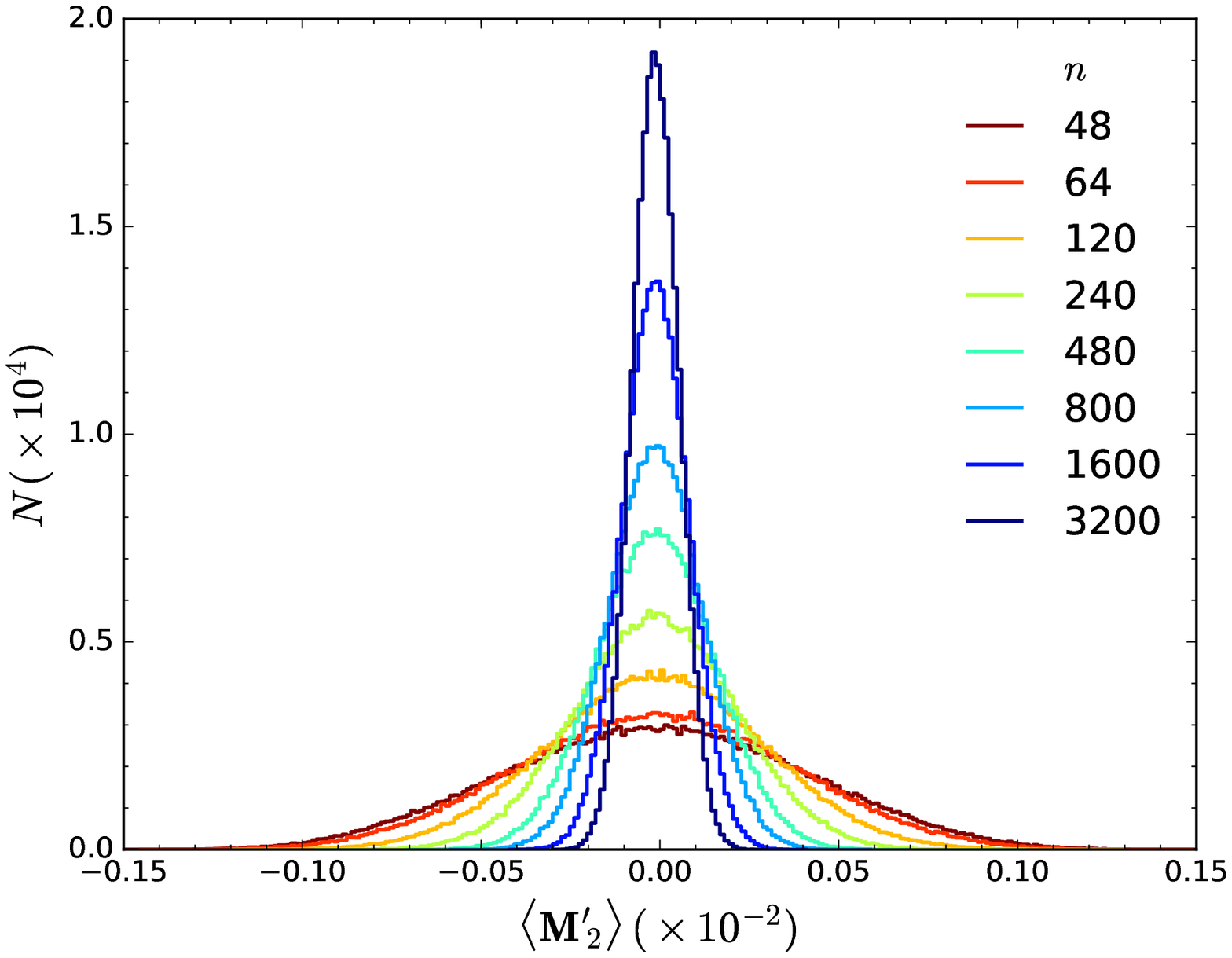}
\caption{Distributions of the moments of the kernel $M_{0}$ and ${\bf M}_{1}$ as given by
Eqs. (18) and (19) ({\it upper plots}) and the moments of the kernel gradient ${\bf M}_{0}^{\prime}$,
${\bf M}_{1}^{\prime}$, and ${\bf M}_{2}^{\prime}$ as given by Eqs. (20) and (21)
({\it middle and bottom plots}) for the random particle distribution, using the Wendland C$^{4}$
function with increasing number of neighbors from $n=48$ (red curves) to 3200 (dark-blue curves).
$\langle {\bf M}_{1}\rangle$ and $\langle {\bf M}_{0}^{\prime}\rangle$ are the means of the
components of ${\bf M}_{1}$ and ${\bf M}_{0}^{\prime}$, while $\langle {\bf M}_{1}^{\prime}\rangle$
and $\langle {\bf M}_{2}^{\prime}\rangle$ correspond to the mean of the elements of matrices
${\bf M}_{1}^{\prime}$ and ${\bf M}_{2}^{\prime}$, respectively.
\label{fig:f3}}
\end{figure}

\clearpage

\begin{figure}
\figurenum{4}
\epsscale{1.0}
\includegraphics[width=1.0\textwidth,angle=-90]{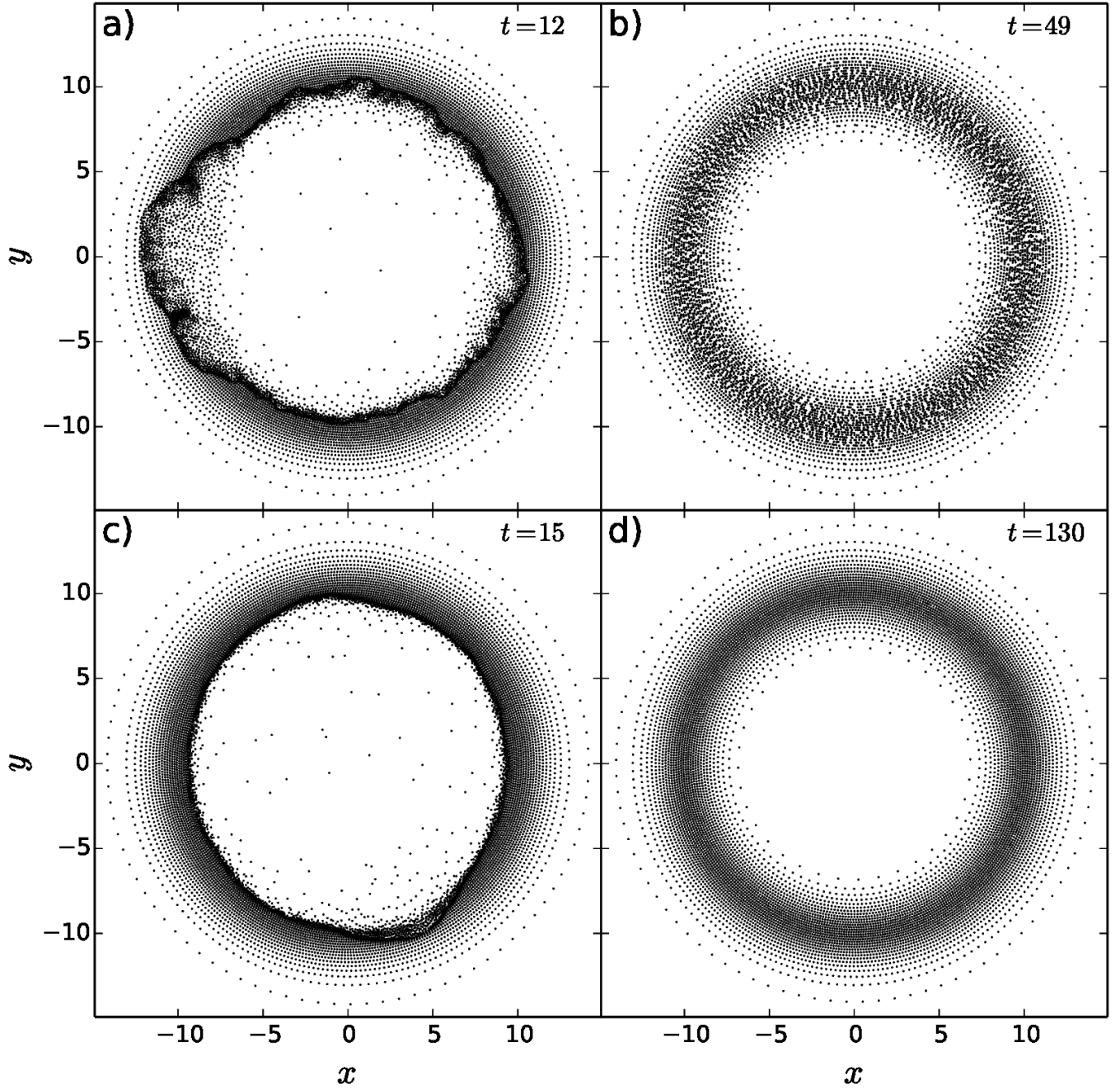}
\caption{Particle positions for the Keplerian ring test using four different SPH
schemes: (a) standard GADGET-2 with the cubic $B$-spline and $n=12$ neighbors;
(b) cubic $B$-spline with $n=12$ and \cite{Hu14} scheme for the artificial viscosity;
(c) standard artificial viscosity and Wendland C$^{4}$ function with $n=120$; and
(d) modified GADGET-2 code with \cite{Hu14} artificial viscosity and Wendland C$^{4}$ 
function with $n=120$. Only for this last scheme the ring preserves its initial
configuration and remains stable.
\label{fig:f4}}
\end{figure}

\clearpage

\begin{figure}
\figurenum{5}
\epsscale{0.8}
\plotone{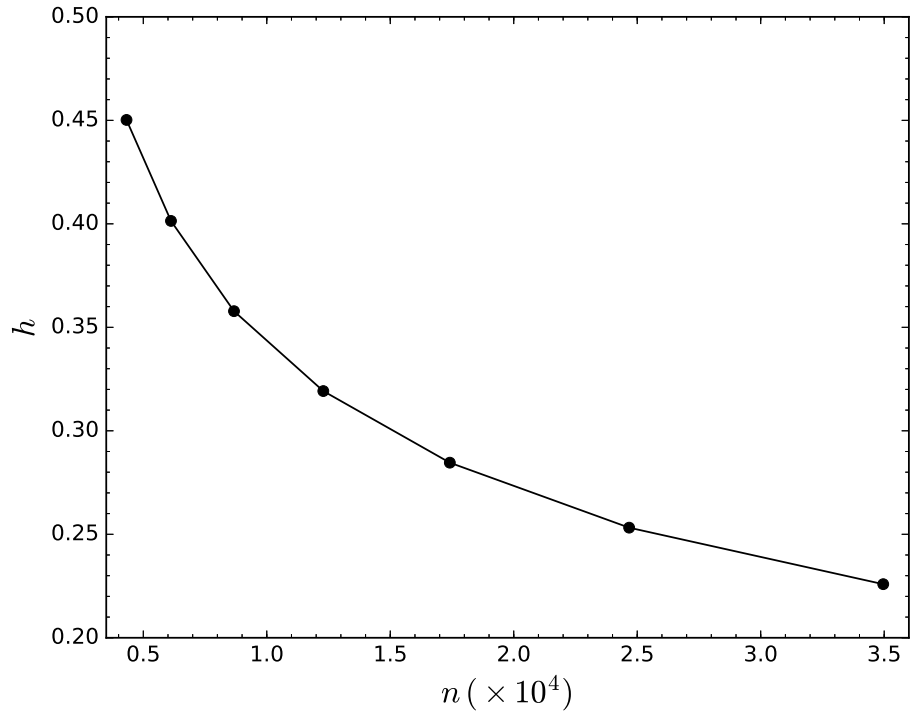}
\caption{Variation of the smoothing length, $h$, with the number of neighbors, $n$,
according to the scaling relation $h\approx 7.23n^{-0.33}$. The dots on the curve
mark the values of $n$ used for the collapse calculations of Table 2 with the C$^{4}$
Wendland function. 
\label{fig:f5}}
\end{figure}

\clearpage

\begin{figure}
\figurenum{6}
\epsscale{1.0}
\plotone{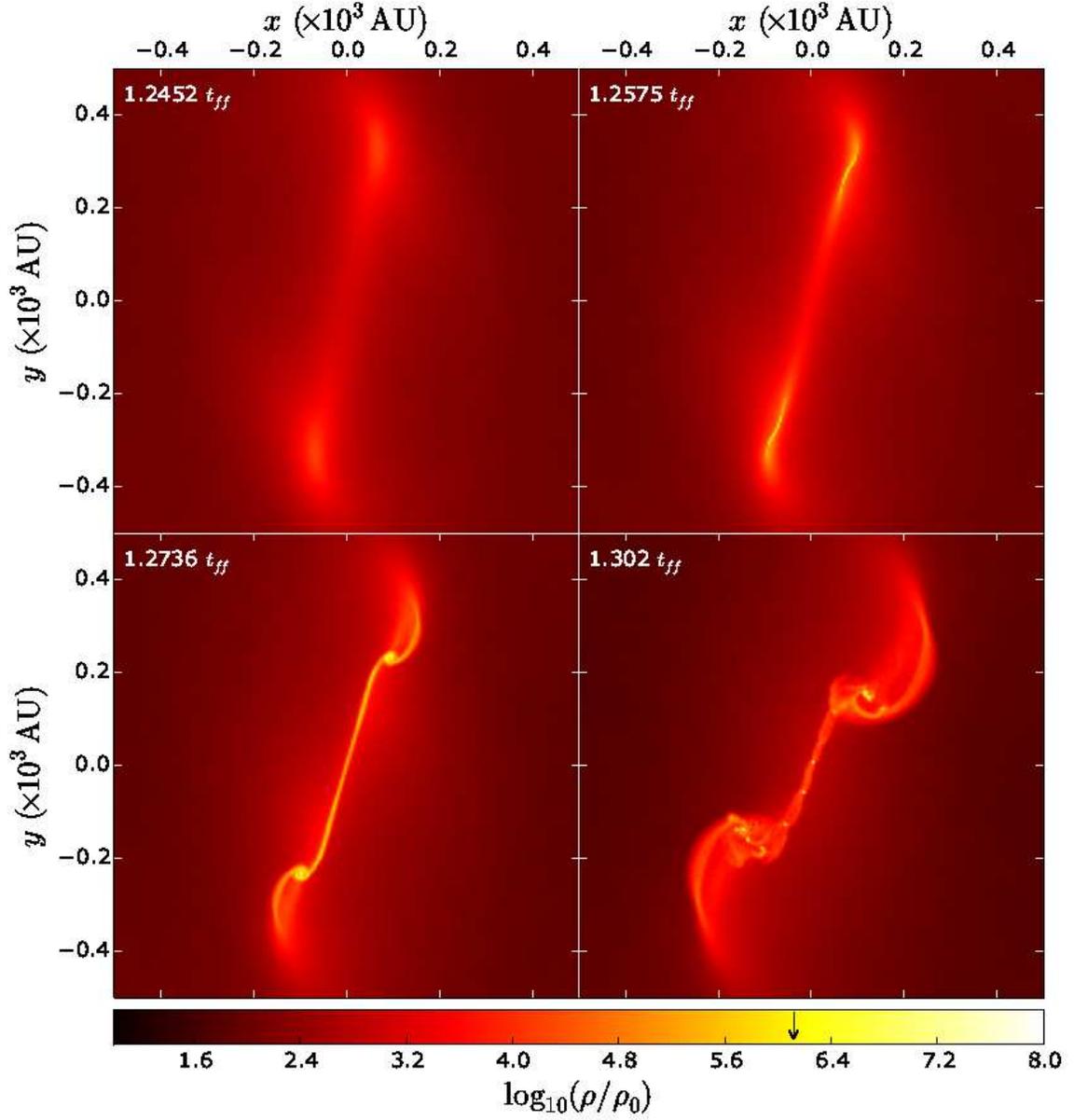}
\caption{Density maps of the cloud center in the equatorial plane for model U4C. Four
different times are shown: $t=1.2452t_{\rm ff}$ ($\rho _{\rm max}\approx 10^{4.22}\rho _{0}$),
$t=1.2575t_{\rm ff}$ ($\rho _{\rm max}\approx 10^{6.97}\rho _{0}$),
$t=1.2736t_{\rm ff}$ ($\rho _{\rm max}\approx 10^{8.91}\rho _{0}$), and
$t=1.302t_{\rm ff}$ ($\rho _{\rm max}\approx 10^{9.74}\rho _{0}$). The color bar at the
bottom shows the logarithm of the density normalized to the initial value $\rho _{0}$
and the vertical arrow marks the critical density beyond which the collapse becomes
adiabatic.
\label{fig:f6}}
\end{figure}

\clearpage

\begin{figure}
\figurenum{7}
\epsscale{1.0}
\plotone{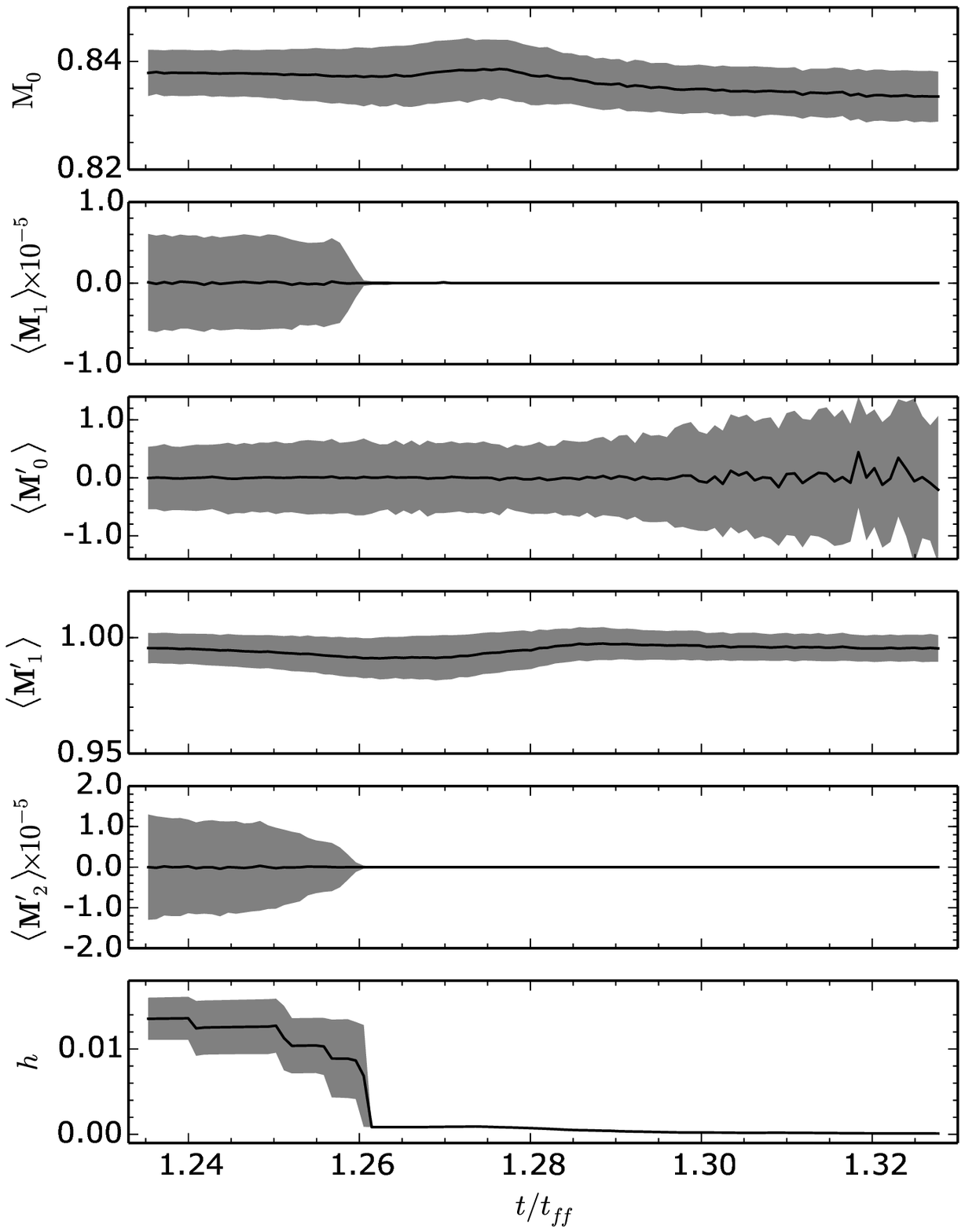}
\caption{Time evolution of the discrete first moments (18)--(21) and smoothing
length for the last stages of collapse of model U4C. The solid line in each plot
represents the maximum of the distribution for each quantity where most particles lie,
while the gray strips around the maximum correspond to distributions composed of 
much lower numbers of particles. Bracketed quantities have the same meaning as in
Figures 2 and 3. 
\label{fig:f7}}
\end{figure}

\clearpage

\begin{figure}
\figurenum{8}
\epsscale{0.8}
\plotone{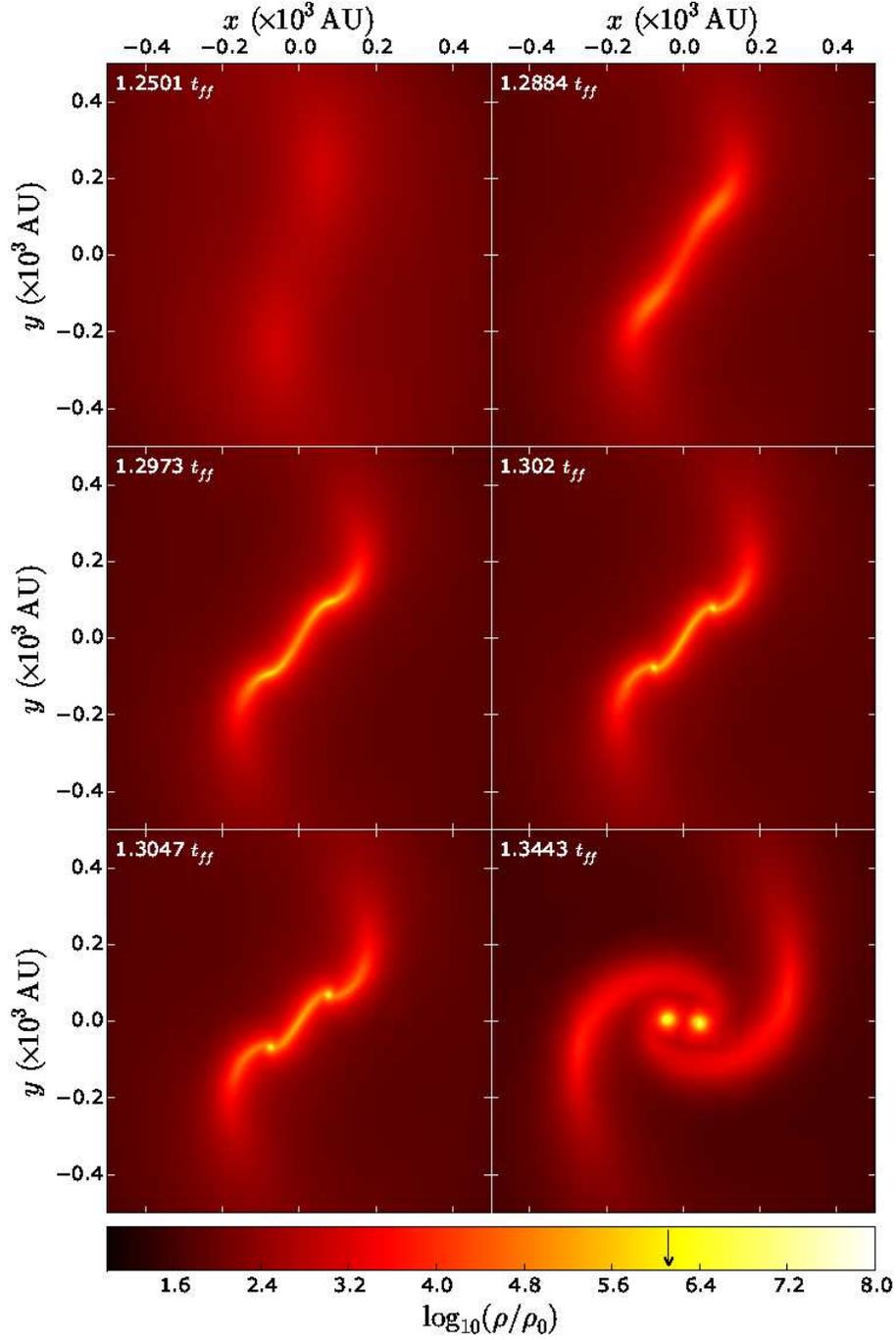}
\caption{Density maps of the cloud center in the equatorial plane for model U2W. Six
different times are shown: $t=1.2501t_{\rm ff}$ ($\rho _{\rm max}\approx 10^{3.05}\rho _{0}$),
$t=1.2884t_{\rm ff}$ ($\rho _{\rm max}\approx 10^{4.84}\rho _{0}$),
$t=1.2973t_{\rm ff}$ ($\rho _{\rm max}\approx 10^{6.75}\rho _{0}$),
$t=1.302t_{\rm ff}$ ($\rho _{\rm max}\approx 10^{8.43}\rho _{0}$), 
$t=1.3047t_{\rm ff}$ ($\rho _{\rm max}\approx 10^{8.64}\rho _{0}$), and 
$t=1.3443t_{\rm ff}$ ($\rho _{\rm max}\approx 10^{9.47}\rho _{0}$). The color bar at the
bottom shows the logarithm of the density normalized to the initial value $\rho _{0}$
and the vertical arrow marks the critical density beyond which the collapse becomes
adiabatic.
\label{fig:f8}}
\end{figure}

\clearpage

\begin{figure}
\figurenum{9}
\epsscale{0.8}
\plotone{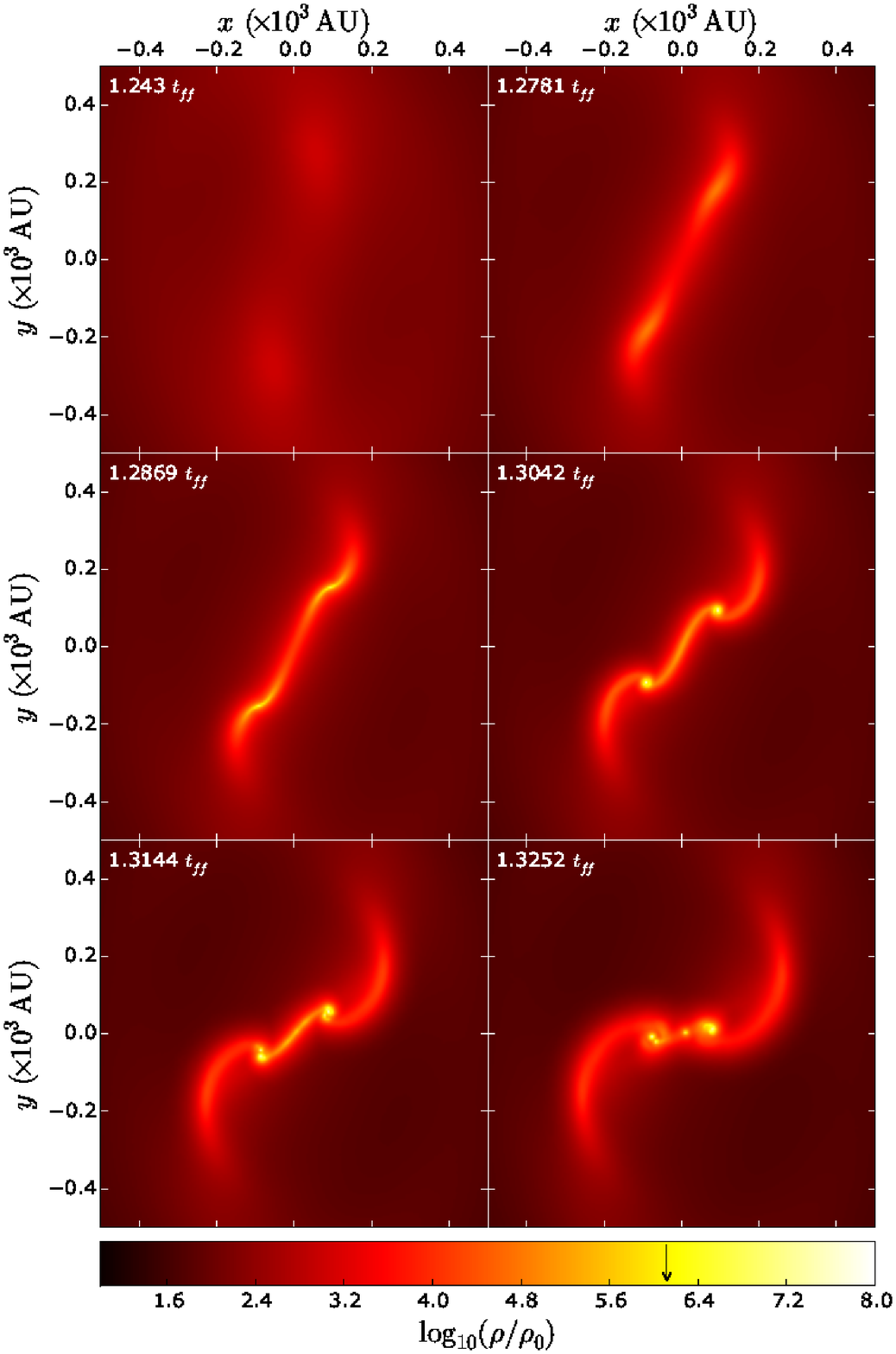}
\caption{Density maps of the cloud center in the equatorial plane for model U4W. Six
different times are shown: $t=1.243t_{\rm ff}$ ($\rho _{\rm max}\approx 10^{3.0}\rho _{0}$),
$t=1.2781t_{\rm ff}$ ($\rho _{\rm max}\approx 10^{4.84}\rho _{0}$),
$t=1.2869t_{\rm ff}$ ($\rho _{\rm max}\approx 10^{6.75}\rho _{0}$),
$t=1.3042t_{\rm ff}$ ($\rho _{\rm max}\approx 10^{8.43}\rho _{0}$),
$t=1.3144t_{\rm ff}$ ($\rho _{\rm max}\approx 10^{8.62}\rho _{0}$), and
$t=1.3252t_{\rm ff}$ ($\rho _{\rm max}\approx 10^{9.46}\rho _{0}$). The color bar at the
bottom shows the logarithm of the density normalized to the initial value $\rho _{0}$
and the vertical arrow marks the critical density beyond which the collapse becomes
adiabatic.
\label{fig:f9}}
\end{figure}

\clearpage

\begin{figure}
\figurenum{10}
\epsscale{1.0}
\plotone{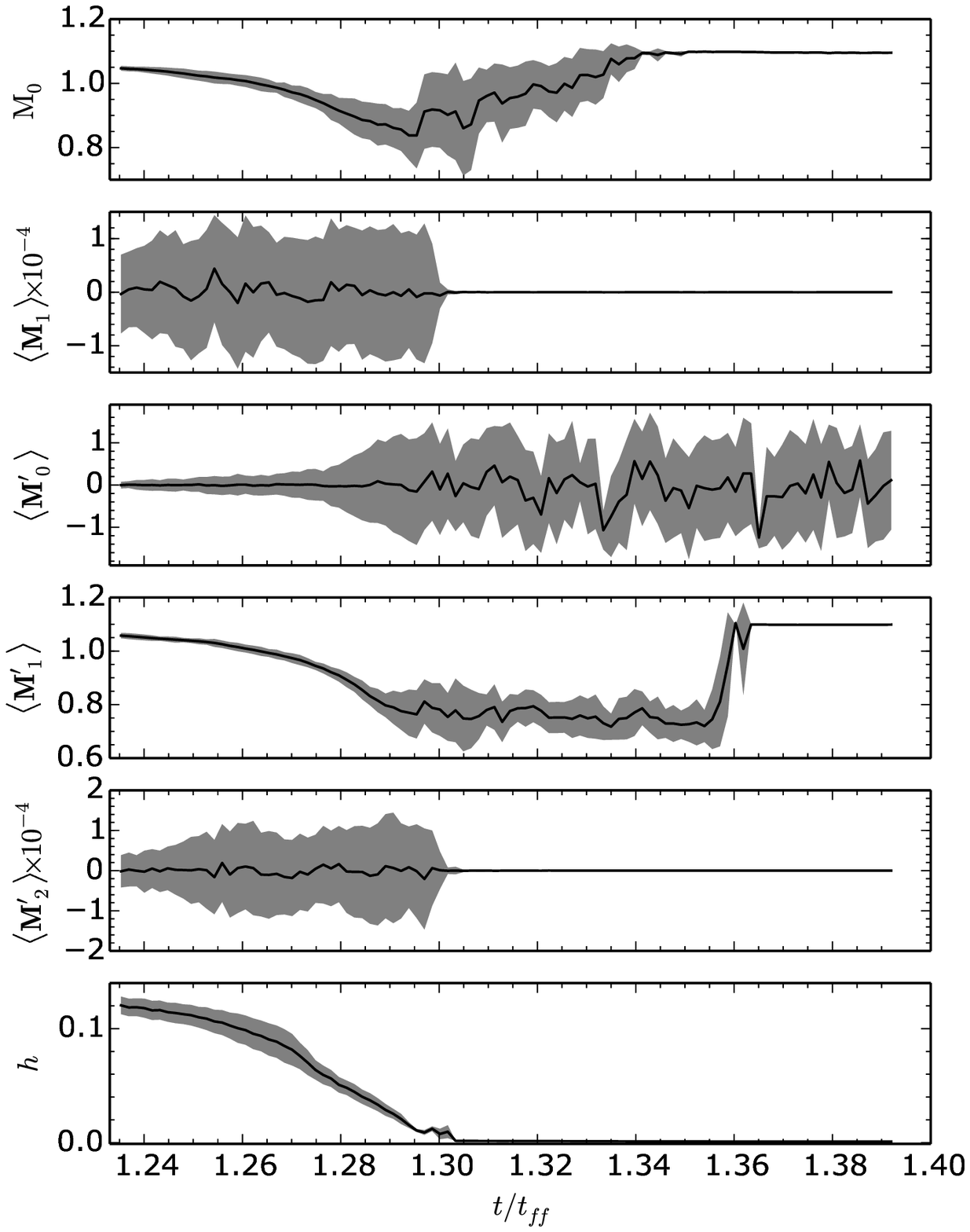}
\caption{Time evolution of the discrete first moments (18)--(21) and smoothing
length for the last stages of collapse of model U2W. The solid line in each plot
represents the maximum of the distribution for each quantity where most particles lie,
while the gray strips around the maximum correspond to distributions composed of 
much lower numbers of particles. Bracketed quantities have the same meaning as
in Figures 2 and 3.
\label{fig:f10}}
\end{figure}

\clearpage

\begin{figure}
\figurenum{11}
\epsscale{1.0}
\plotone{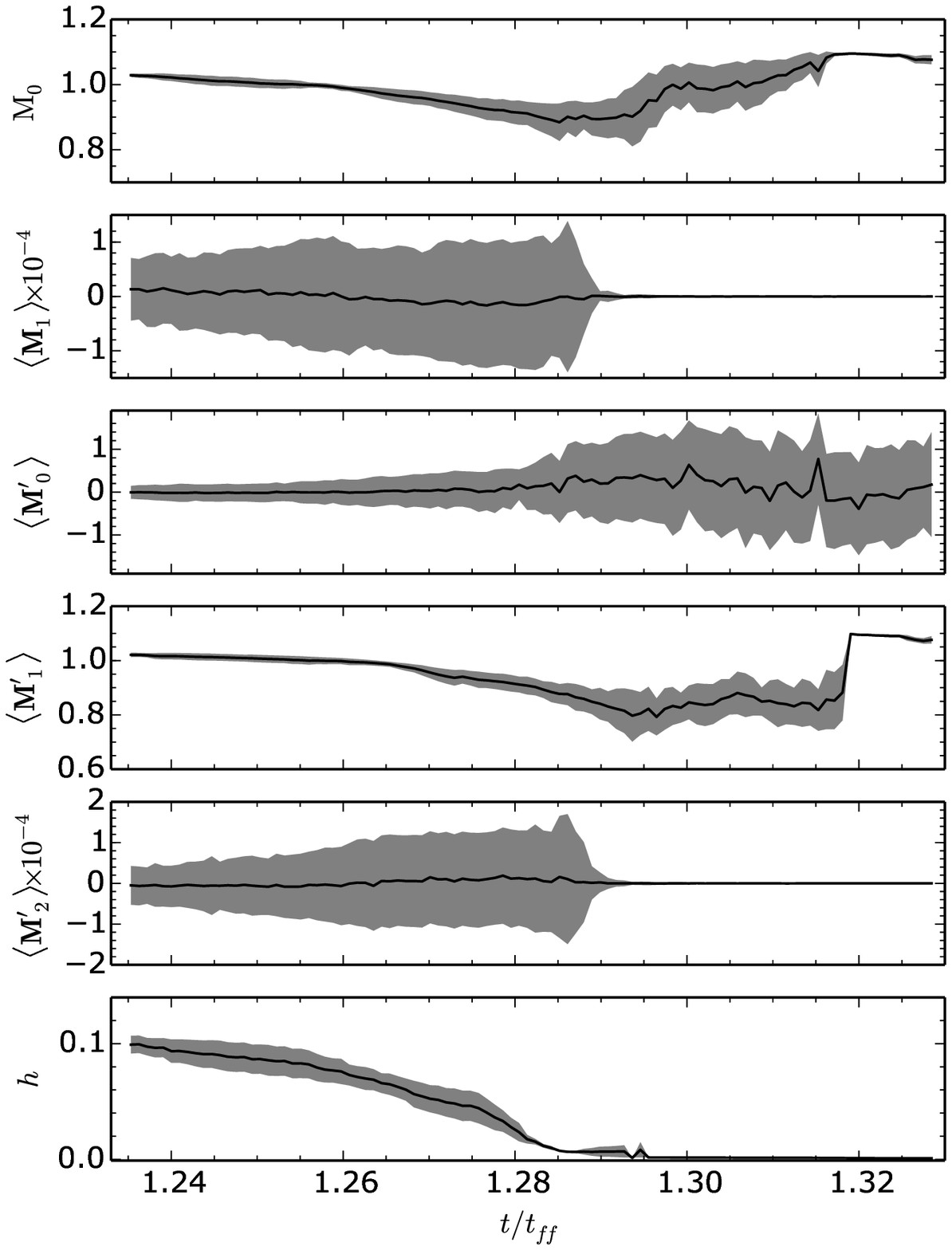}
\caption{Time evolution of the discrete first moments (18)--(21) and smoothing
length for the last stages of collapse of model U4W. The solid line in each plot
represents the maximum of the distribution for each quantity where most particles lie,
while the gray strips around the maximum correspond to distributions composed of much 
lower numbers of particles. Bracketed quantities have the same meaning as in
Figures 2 and 3.
\label{fig:f11}}
\end{figure}

\clearpage

\begin{figure}
\figurenum{12}
\epsscale{1.0}
\plotone{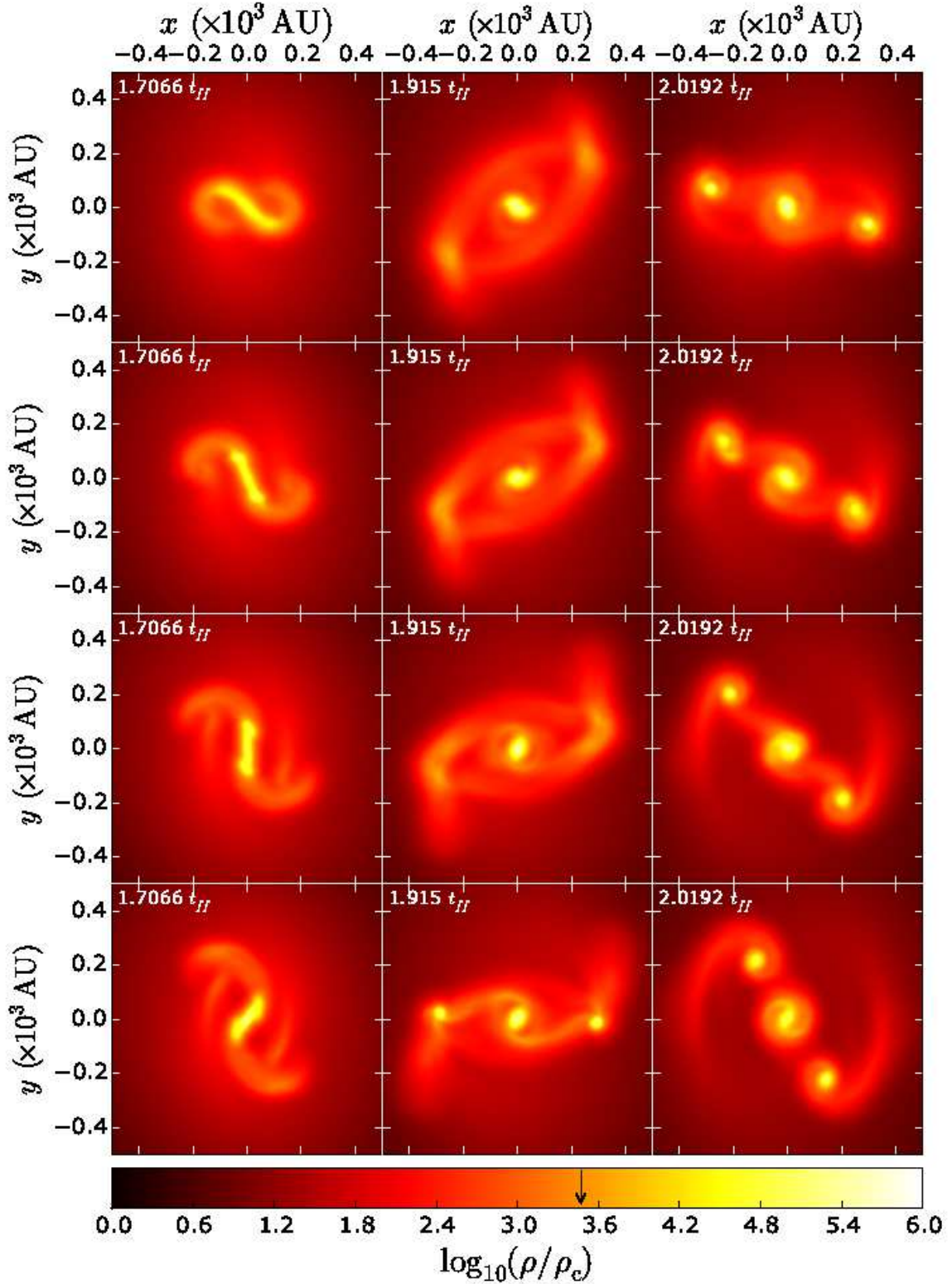}
\caption{Density maps of the cloud center in the equatorial plane for models G3W (first
row, $N=1200000$), G4W (second row, $N=2400000$), G5W (third row, $N=4800000$), and
G6W (last row, $N=9600000$). Three different times are displayed for each model from 
left to right showing the intermediate and final stages of collapse. At $1.7066t_{\rm ff}$,
$\rho _{\rm max}\approx 10^{4.70}\rho _{\rm c}$ (G3W), $\approx 10^{5.07}\rho _{\rm c}$ (G4W),
$\approx 10^{5.07}\rho _{\rm c}$ (G5W), and $\approx 10^{5.35}\rho _{\rm c}$ (G6W); at $1.915t_{\rm ff}$,
$\rho _{\rm max}\approx 10^{5.42}\rho _{\rm c}$ (G3W), $\approx 10^{5.41}\rho _{\rm c}$ (G4W),
$\approx 10^{5.57}\rho _{\rm c}$ (G5W), and $\approx 10^{5.54}\rho _{\rm c}$ (G6W); and at $2.0192_{\rm ff}$,
$\rho _{\rm max}\approx 10^{5.58}\rho _{\rm c}$ (G3W), $\approx 10^{5.39}\rho _{\rm c}$ (G4W),
$\approx 10^{5.44}\rho _{\rm c}$ (G5W), and $\approx 10^{5.32}\rho _{\rm c}$ (G6W), where 
$\rho _{\rm c}=1.7\times 10^{-17}$ g cm$^{-3}$ is the initial central density. The color
bar at the bottom shows the logarithm of the density normalized to $\rho _{\rm c}$ and
the vertical arrow marks the critical density beyond which the collapse becomes adiabatic.
\label{fig:f12}}
\end{figure}

\clearpage

\begin{figure}
\figurenum{13}
\epsscale{1.0}
\plotone{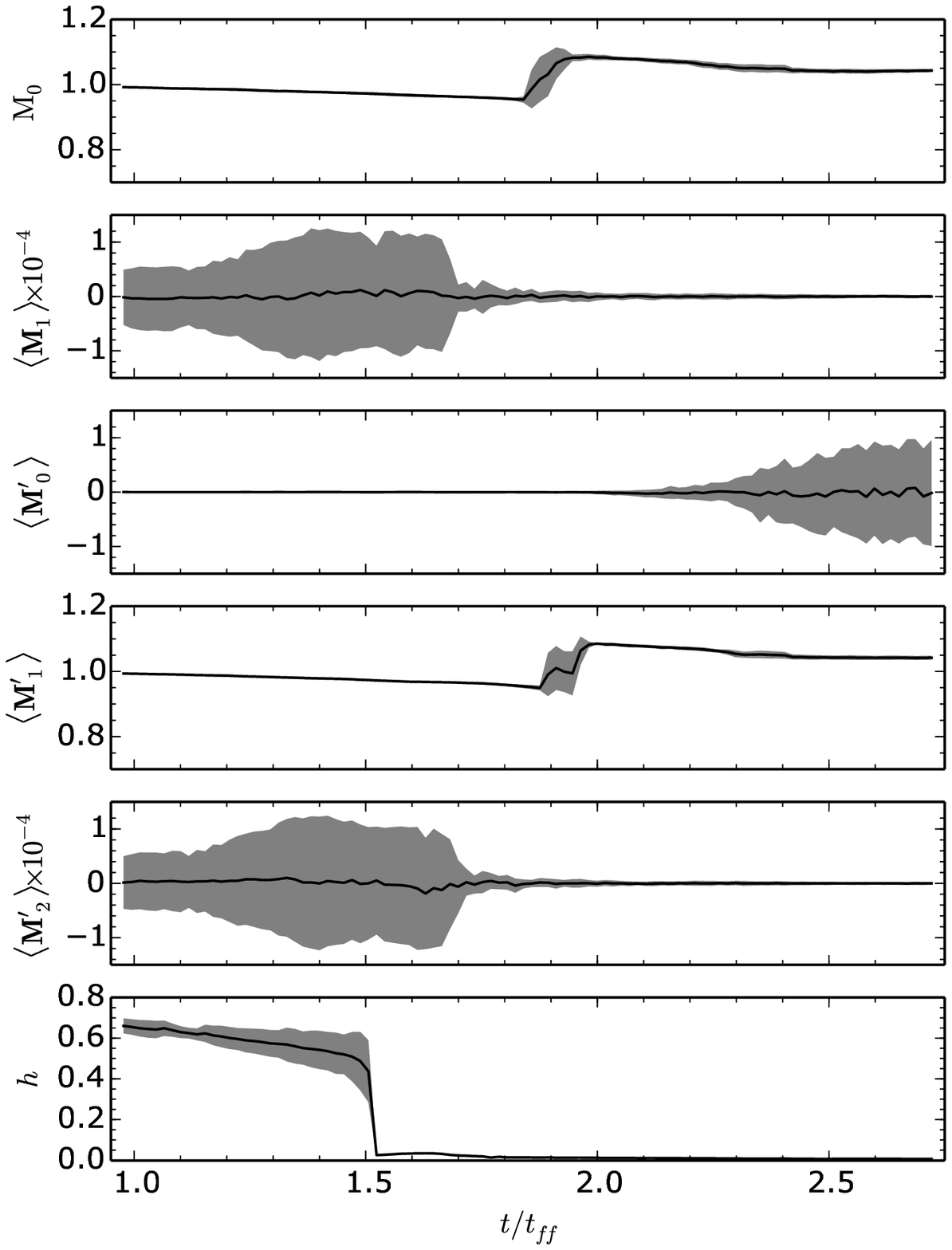}
\caption{Time evolution of the discrete first moments (18)--(21) and smoothing
length for the last stages of collapse of model G3W. The solid line in each plot
represents the maximum of the distribution for each quantity where most particles lie,
while the gray strips around the maximum correspond to distributions composed of much 
lower numbers of particles. Bracketed quantities have the same meaning as in
Figures 2 and 3.
\label{fig:f13}}
\end{figure}

\clearpage

\begin{figure}
\figurenum{14}
\epsscale{1.0}
\plotone{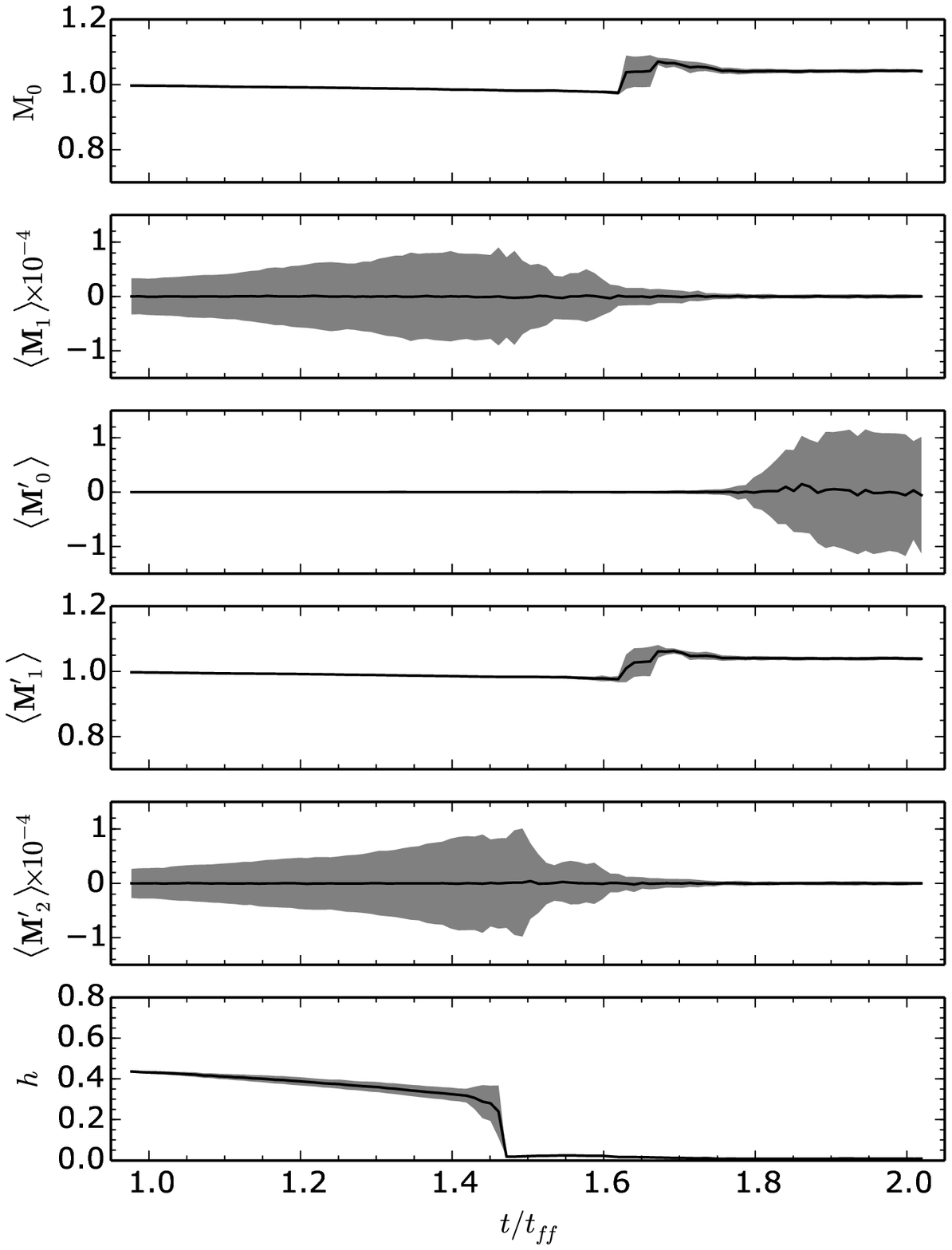}
\caption{Time evolution of the discrete first moments (18)--(21) and smoothing
length for the last stages of collapse of model G6W. The solid line in each plot
represents the maximum of the distribution for each quantity where most particles lie,
while the gray strips around the maximum correspond to distributions composed of much 
lower numbers of particles. Bracketed quantities have the same meaning as in
Figures 2 and 3.
\label{fig:f14}}
\end{figure}

\clearpage

\begin{figure}
\figurenum{15}
\epsscale{0.8}
\plotone{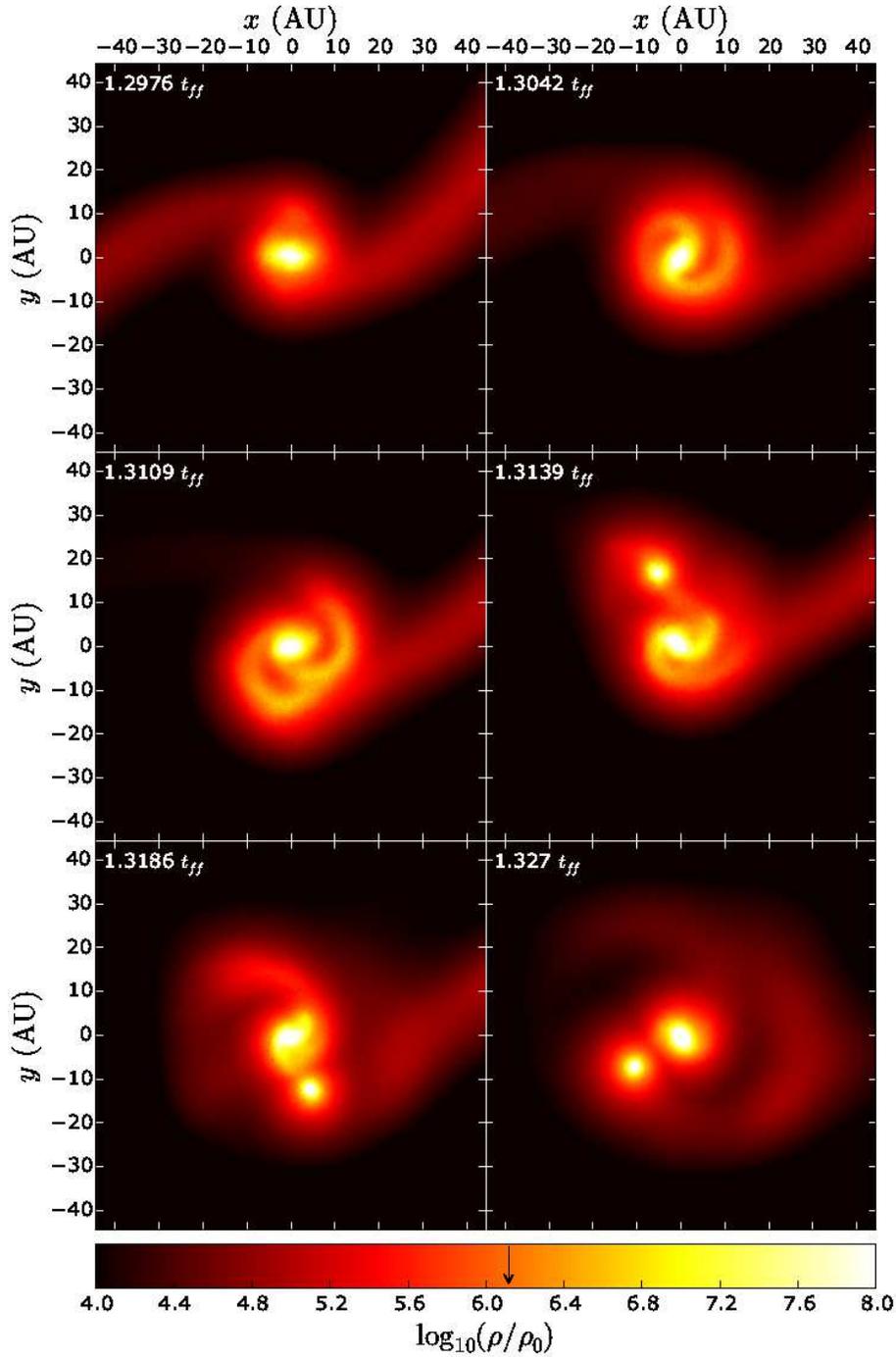}
\caption{Enlarged density maps in the equatorial plane of one of the binary protostars
that formed in model U4W. Six different times are shown: $t=1.2976t_{\rm ff}$
($\rho _{\rm max}\approx 10^{8.23}\rho _{0}$), $t=1.3042t_{\rm ff}$
($\rho _{\rm max}\approx 10^{8.42}\rho _{0}$), $t=1.3109t_{\rm ff}$
($\rho _{\rm max}\approx 10^{8.53}\rho _{0}$), $t=1.3139t_{\rm ff}$
($\rho _{\rm max}\approx 10^{8.64}\rho _{0}$), $t=1.3186t_{\rm ff}$
($\rho _{\rm max}\approx 10^{8.89}\rho _{0}$), $t=1.3270t_{\rm ff}$
($\rho _{\rm max}\approx 10^{9.77}\rho _{0}$). A close binary of maximum orbital
separation $\approx 20$ AU is formed by fragmentation of the gravitationally
unstable disk around the primary. The color bar at the bottom shows the logarithm of
the density normalized to the initial value $\rho _{0}$ and the vertical arrow marks
the critical density beyond which the collapse becomes adiabatic. 
\label{fig:f15}}
\end{figure}


\end{document}